%% file: main.tex
\documentclass[runningheads]{llncs}
\usepackage[T1]{fontenc}
\usepackage{graphicx}
\usepackage{subfig}

\newcounter{hypcounter}
\renewcommand{\thehypcounter}{\arabic{hypcounter}}
\newenvironment{hyp}[1]%
{\refstepcounter{hypcounter}\par\addvspace{\medskipamount}\noindent\textbf{Hypothesis \thehypcounter\ (H\thehypcounter):}\enskip}%
{\par\addvspace{\medskipamount}}
\usepackage{multirow}
\usepackage{hyperref}
\usepackage{cleveref}
\usepackage{wrapfig}
\usepackage{caption}
\usepackage{soul}
\usepackage{wrapfig}

\renewcommand{\paragraph}[1]{\textbf{#1:}}
\crefname{figure}{Fig.}{Figs.}
\Crefname{figure}{Fig.}{Figs.}

\newcommand{\chiSqrAPA}[4]{${ \chi}^2\left(#1, N\!=\!#2\right)\!=\!#3$, $p#4$}

\newcommand{\tTestAPA}[3]{$t\left(#1\right)\!=\!#2$, $p#3$}

\newcommand{\MeanSdAPA}[2]{(M$=$#1, \allowbreak\ SD$=$#2)}

\begin{document}
\title{Evaluating Line Chart Strategies for Mitigating Density of Temporal Data: The Impact on Trend, Prediction, and Decision-Making}
\titlerunning{Line Chart Strategies for Mitigating Density of Temporal Data}
\author{Rifat Ara Proma\inst{1} \and
Ghulam Jilani Quadri\inst{2} \and
Paul Rosen\inst{1}}
\authorrunning{R. Proma et al.}
\institute{University of Utah, Salt Lake City, UT 84112, USA \\
\email{rifat.proma@utah.edu} and \email{paul.rosen@utah.edu}\\
\and
University of Oklahoma, Norman, OK 73019, USA \\
\email{quadri@ou.edu}\\
}

\maketitle              %

\input{sec.abstract}

\input{sec.intro}

\input{sec.lit-review.tex}

\input{sec.design}

\input{sec.main_study}

\input{sec.tasks}

\input{sec.discussion_future_work}

\input{sec.conclusion}

\bibliographystyle{splncs04}
\bibliography{main}

\end{document}

%% file: sec.abstract.tex
\begin{abstract}

Overplotted line charts can obscure trends in temporal data and hinder prediction. We conduct a user study comparing three alternatives—aggregated, trellis, and spiral line charts against standard line charts on tasks involving trend identification, making predictions, and decision‑making. We found aggregated charts performed similarly to standard charts and support more accurate trend recognition and prediction; trellis and spiral charts generally lag. We also examined the impact on decision-making via a trust game. The results showed similar trust in standard and aggregated charts, varied trust in spiral charts, and a lean toward distrust in trellis charts. These findings provide guidance for practitioners choosing visualization strategies for dense temporal data.

\keywords{Temporal data, line charts, data aggregation, trellis charts, spiral charts.}
\end{abstract}

%% file: sec.intro.tex
\section{Introduction}

\begin{wrapfigure}{r}{0.295\linewidth}

    \centering
    \includegraphics[width=0.975\linewidth]{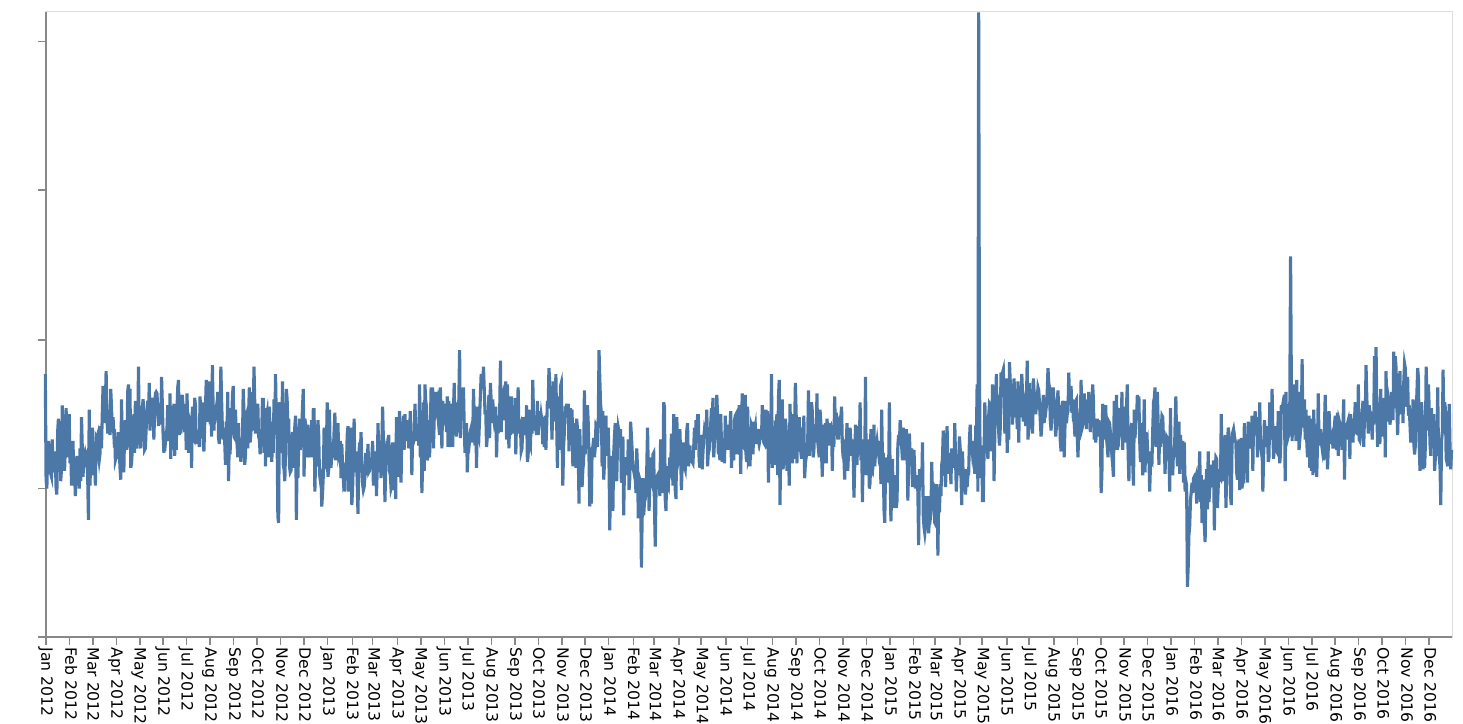}

    \caption{Line chart showing daily crime incidents in Baltimore.} 
    
\label{fig:line-baltimore}
\end{wrapfigure}

Temporal data analysis is critical in domains ranging from finance to healthcare. Although many chart types can display time‑varying data, line charts are 
widely used, generally outperforming bar charts and scatterplots for detecting nonlinear trends~\cite{best2007perception}. However, when there are many data points, line charts can become dense and cluttered (Fig.~\ref{fig:line-baltimore}), which impairs judgment accuracy~\cite{ryan2018glance} and makes it hard to see patterns or make informed decisions~\cite{aigner2007visual}. Interactive techniques like panning and zooming mitigate overplotting~\cite{walker2015timenotes} but can cause loss of context~\cite{glueck2013model}.
We focus on static alternatives that modify the chart layout or representation. 

Building on Brehmer et al.’s timeline design space~\cite{brehmer2016timelines}, we consider trellis charts (\cref{fig:trellis-climate}) that segment a time series into stacked panels to exploit vertical space and highlight periodic behavior~\cite{schreck2007trajectory} and spiral charts (\cref{fig:spiral-climate}) that map time onto a coil to compactly reveal cycles~\cite{draper2009survey,weber2001visualizing}. We also consider smoothing by data aggregation (\cref{fig:aggregated-climate}), which is not part of Brehmer et al.'s design space. Aggregation reduces the number of samples and clarifies the signal~\cite{aigner2007visual,rosen2020linesmooth}. However, while all of these techniques mitigate the effects of density, their impact on analysis and decision-making has not been systematically studied.

We therefore conduct a human‑subjects study with 72 Prolific participants to evaluate aggregated, trellis, and spiral line charts against a standard line chart with all data points. Although such charts can be cluttered, they provide 
the most detailed view of the data, making them a useful baseline. This allows us to assess whether reduced-density alternatives can match or exceed the standard chart in interpretability, accuracy, and trust. Guided by Ciccione and Dehaene’s multistep model of chart comprehension~\cite{ciccione2021can}, we examine how each alternative, relative to standard line charts, supports: 1) identifying trends, 2) producing consistent predictions about future values, and 3) making decisions with trust in the visualization. The findings inform guidelines for choosing visualization strategies for dense temporal data.

%% file: sec.lit-review.tex
\section{Prior Work}

Line charts are widely used for representing temporal data in scientific publications and public communication~\cite{borkin2013makes}. Because they are so common, a substantial body of research has examined how well they support tasks, including correlations~\cite{harrison2014ranking}, trend detection~\cite {correll2017regression}, outlier identification~\cite{albers2014task}, and data distributions~\cite{gogolou2018comparing}. Waldner et al.~\cite{waldner2019comparison} found that users generally prefer linear charts for low-level tasks. However, factors such as aspect ratio and color or positional encoding can influence perception in linear charts~\cite{heer2009sizing}. Moreover, dense time series lead to overplotting and visual clutter, motivating alternative designs.

Studies suggest tackling density by summarizing the data. Aggregation and smoothing decrease the number of samples and clarify the overall signal in noisy series~\cite{aigner2007visual,rosen2020linesmooth}. Another branch of research recommends splitting the data into coordinated panels such as trellis charts, horizon graphs, and other small‑multiple layouts to enable side‑by‑side comparison and mitigate overplotting~\cite{becker1996visual,javed2010graphical,perin2013interactive}. Some studies also suggest radial layouts, including spiral charts, which map time onto a coil to emphasize periodicity while fitting long sequences into a compact space~\cite{hewagamage1999interactive,weber2001visualizing}. Each of these designs offers advantages in particular scenarios but also introduces trade‑offs. For example, radial layouts can distort values or compress parts of the display~\cite{stasko2000focus+}. Empirical comparisons have shown that simple linear charts often yield faster task completion than more complex designs~\cite{waldner2019comparison}, yet the impact of these choices on higher‑level outcomes remains unclear.

Given that overplotted charts are harder to comprehend~\cite{aigner2007visual}, it is worth examining how different density-reduction techniques affect decision-making, which relies heavily on trust. Beyond task accuracy, user trust is essential for effective visual analytics~\cite{sacha2015role,thomas2009challenges}. Researchers have begun to adapt methods from psychology and behavioral economics, such as perceptual‑fluency measures, Likert scales, and trust games, to quantify how much users trust a visualization~\cite{alves2022exploring,elhamdadi2022we,elhamdadi2022using}. We adopt the trust-game approach from~\cite{elhamdadi2022we} to assess how different designs influence decision-making compared to a standard line chart.

Various approaches have been put forth to address the challenge posed by dense line charts. However, prior work has not explored whether aggregating data, trellising, or mapping it onto a spiral affects users’ ability to identify trends, make predictions, or trust the visualization. To our knowledge, no study systematically compares these three density‑reduction strategies against a standard line chart on tasks that involve not only perceptual judgment but also prediction and decision‑making. Our work addresses this gap.

%% file: sec.design.tex
\section{Design Motivation}
\label{sec:design}

\begin{figure}[!t]
    \centering
    \begin{minipage}[b]{0.39\linewidth}
        \centering
        \subfloat[Standard line chart \label{fig:line-climate}]{\includegraphics[width=0.975\linewidth]{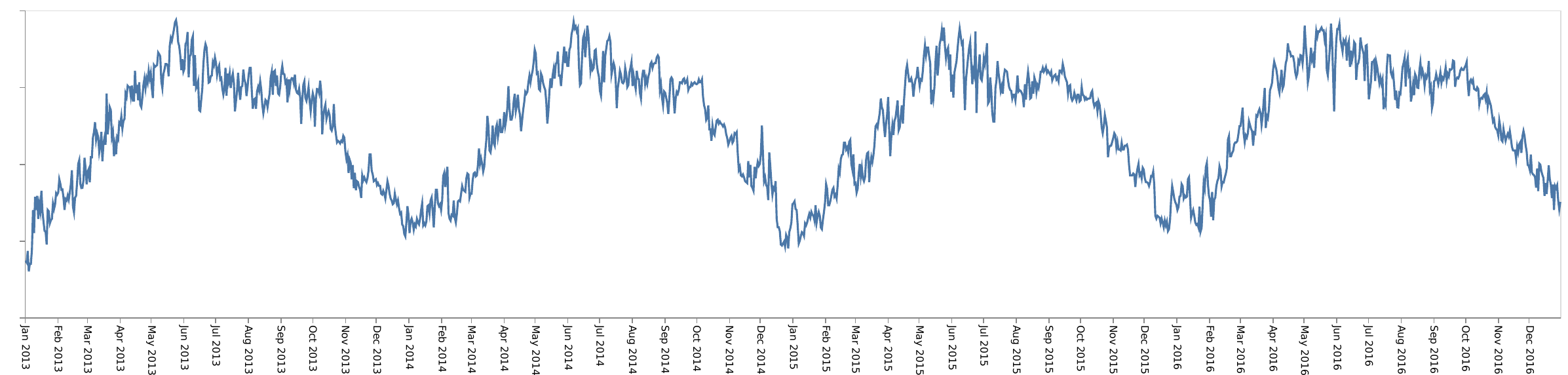}}
        
        \subfloat[Monthly aggregated chart\label{fig:aggregated-climate}]{\includegraphics[width=0.975\linewidth]{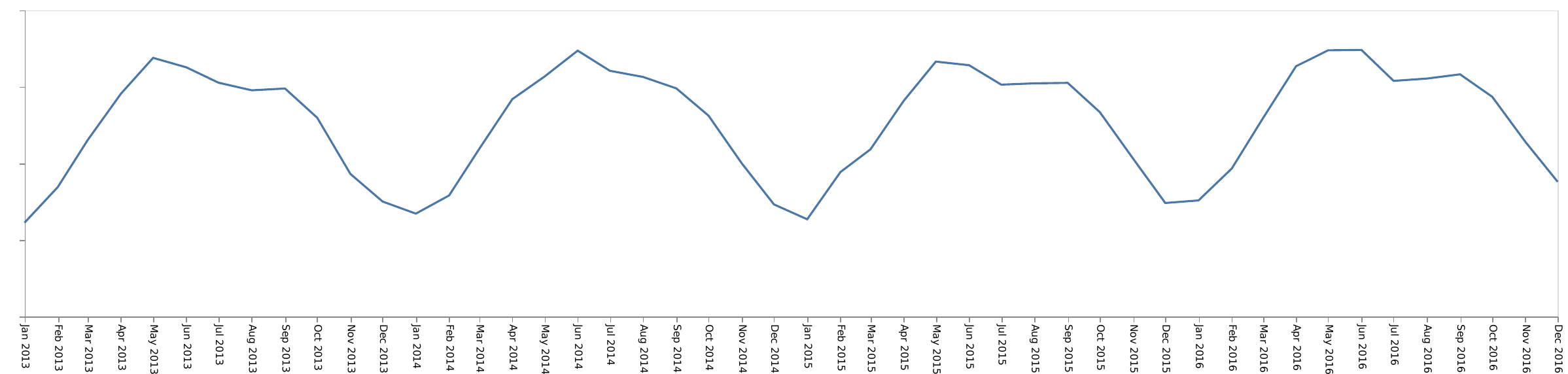}}
    \end{minipage}    
    \hfil
    \begin{minipage}[b]{0.285\linewidth}
        \subfloat[Trellis chart\label{fig:trellis-climate}]{\includegraphics[trim=20pt 0 0 0, clip, width=0.975\linewidth]{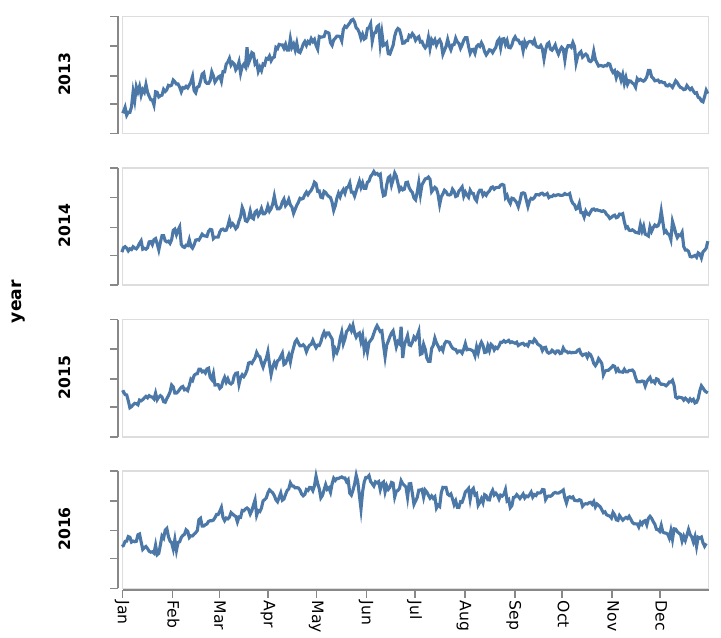}}
    \end{minipage}
    \hfil
    \begin{minipage}[b]{0.26\linewidth}
        \subfloat[Spiral chart\label{fig:spiral-climate}]{\includegraphics[width=0.975\linewidth]{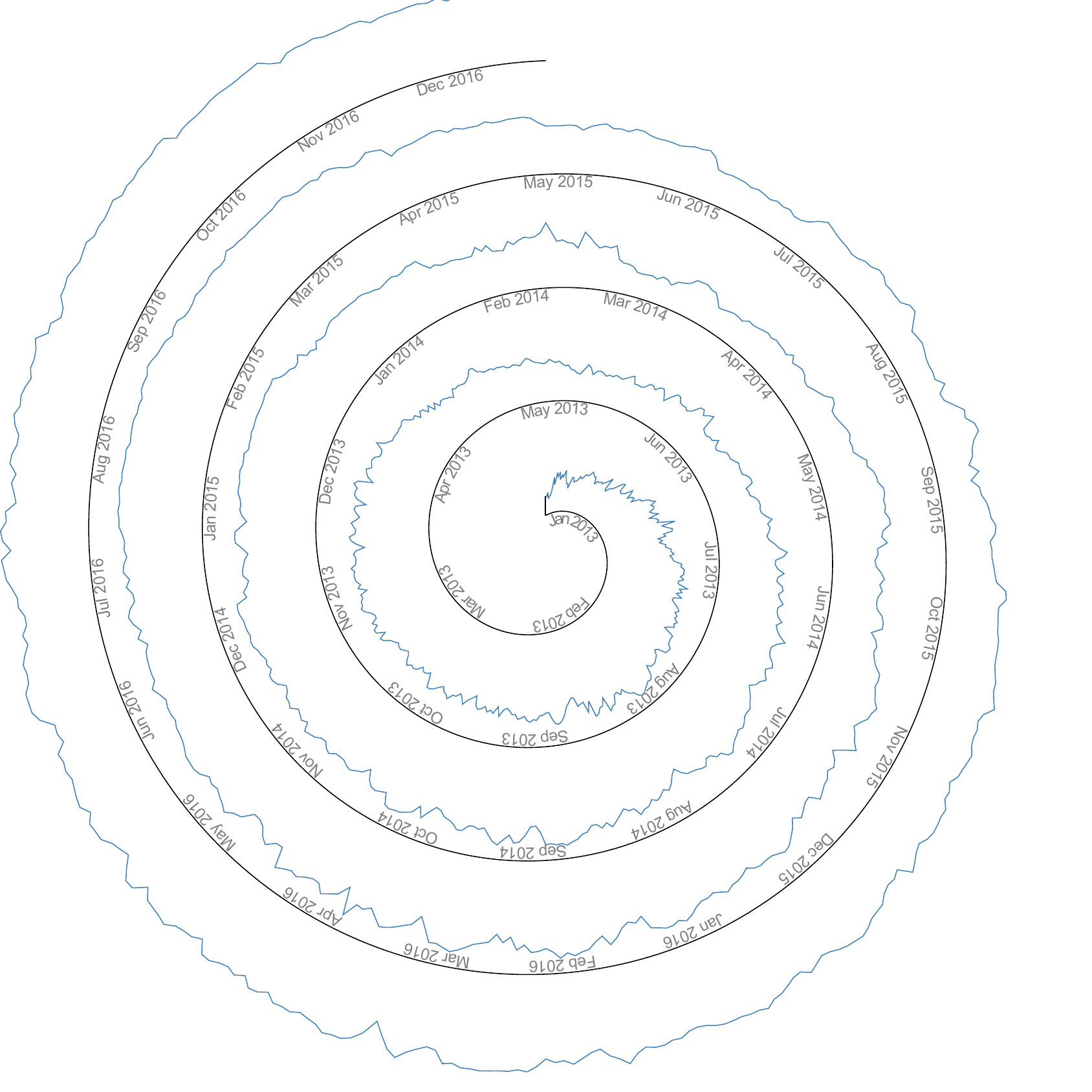}}
    \end{minipage}

\caption{Examples of the line chart visualizations used in the study showing the humidity change data in Delhi from January 2013 to December 2016.}
\label{fig:climate}
\end{figure}

Our study compares four line chart designs to understand how different density-reduction strategies affect users’ ability to interpret temporal data. We focus on line charts because they are ubiquitous and familiar, yet dense series can quickly become cluttered. The variations we test differ in how they balance detail, space utilization, and visual complexity. \Cref{fig:climate} presents examples of all four designs, which form the basis of the user evaluation described in the next section.

\paragraph{Standard Chart}  The baseline is a conventional line chart that plots daily values across the full time span (\Cref{fig:line-climate}).  This format preserves every data point and therefore the fine structure of the series, but long or highly variable sequences can lead to cluttered charts that obscure trends. As we want to compare all the other visualization approaches with this type of chart, we will refer to this chart as the standard chart throughout the paper.

\paragraph{Aggregated Chart}  Inspired by smoothing techniques, this variant groups daily values by month and plots the monthly means (\cref{fig:aggregated-climate}).  Reducing the sampling frequency lowers the number of plotted points, yielding a smoother line that more clearly reveals overarching trends, at the expense of details.

\paragraph{Trellis Chart}  To better utilize vertical screen space and reduce overplotting, we break the series into yearly segments and stack them vertically (\cref{fig:trellis-climate}).  Each panel shows a single year, with axes aligned to facilitate year‑to‑year comparison.  We deliberately avoid color coding to maintain consistency with the visual encoding of other chart types. Splitting by year strikes a compromise between the number of points per panel and the number of panels.

\paragraph{Spiral Chart} Building on radial visualizations, the spiral chart maps time onto a coil (\cref{fig:spiral-climate}): earlier dates begin near the center, and subsequent time steps wind outward. This layout retains daily granularity while fitting a multiyear series into a compact square and highlighting cyclical patterns. By presenting the series as a continuous spiral, it avoids the discontinuities in the trellis chart.

%% file: sec.main_study.tex
\section{Methodology}

\begin{wrapfigure}{r}{0.35\columnwidth}
  \centering
  \includegraphics[width=0.95\linewidth]{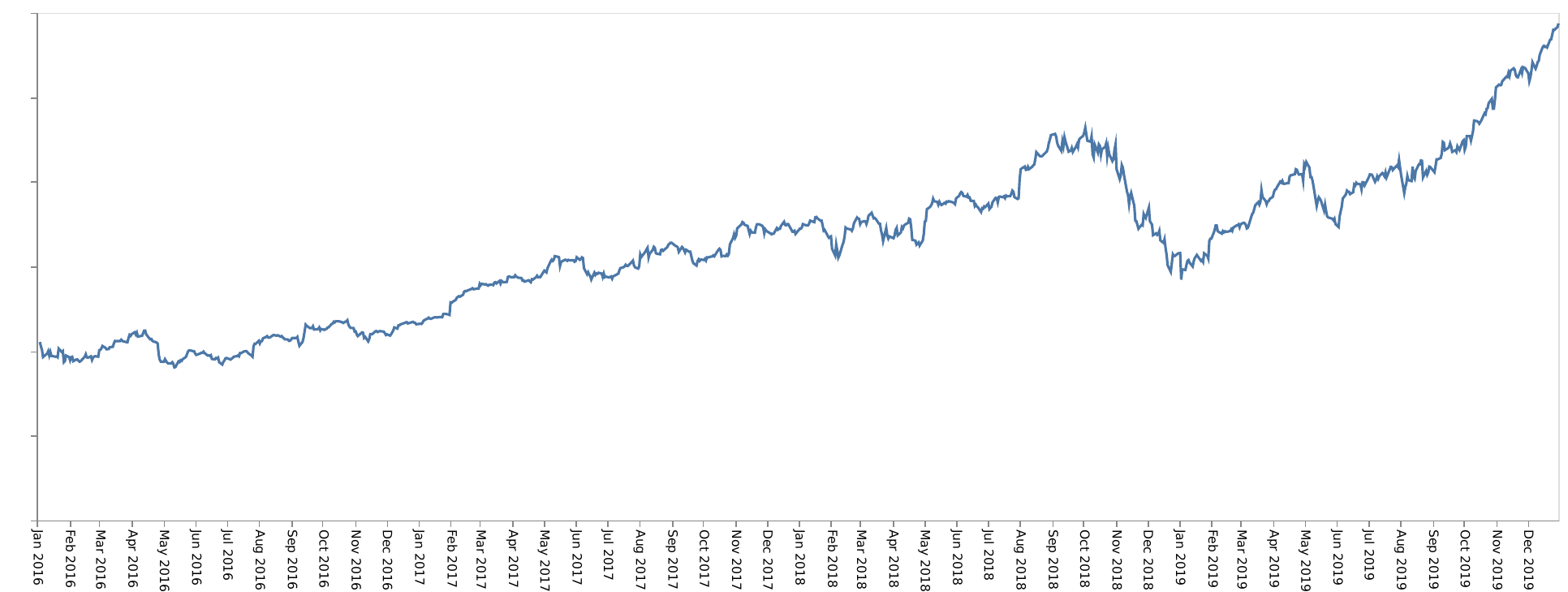}
  \caption{Daily Apple stock price dataset.}
  \label{fig:std-apple-dataset}
\end{wrapfigure}

Our IRB-approved study follows a between-subjects design.  Building on Brehmer et al.’s task typology~\cite{brehmer2013multi} and Ciccione and Dehaene’s model of chart comprehension~\cite{ciccione2021can}, we created tasks to test three abilities: identifying trends, making predictions (with confidence), and making decisions in a trust game.  After a pilot study, we recruited 80 participants via Prolific; 72 remained after attention checks.  Participants were from the US, UK, and Canada, aged 18–64, with a balanced gender distribution and diverse educational backgrounds.

\paragraph{Visualizations}
Participants compared the standard daily line chart with its aggregated, trellis, and spiral variants (Fig.~\ref{fig:climate}).  All charts used a monochrome palette with no gridlines, tick marks, or y‑axis labels for consistency; axes showed only month and year since exact values were not needed for tasks. Aspect ratios were chosen according to Cleveland and McGill’s recommendations~\cite{cleveland1988shape}, and trellis panels were scaled to the number of years shown. Spirals used a fixed square with one coil per year, adjusting the radius to distribute points evenly.

\paragraph{Implementation}
The survey ran on a Flask app with a Firestore database. Charts were rendered using D3.js and Vega-lite. Screenshots of the web application can be found in the \href{https://osf.io/x9zmr/?view_only=c3a7071971e5420ea5f8b99e5af420dd}{supplemental material}.

\paragraph{Dataset}
We selected five publicly available time-series datasets from Kaggle: Apple~\cite{nagadia_2022} (\Cref{fig:std-apple-dataset}) and Tesla~\cite{bozsolik_2020} stock prices, Bitcoin prices~\cite{zielak_2021}, Baltimore crime incidents~\cite{dane_2017} (\cref{fig:line-baltimore}), and humidity in Delhi~\cite{sevgisarac_2021} (\cref{fig:climate}).  Each was truncated to three years and anonymized in the survey to minimize bias. Charts for all datasets are provided in the \href{https://osf.io/x9zmr/?view_only=c3a7071971e5420ea5f8b99e5af420dd}{supplemental material}.

\paragraph{Procedure}
After informed consent and a demographic questionnaire, participants received a brief tutorial on recognizing upward, downward, repeating (i.e., periodic), and constant trends, with practice questions covering trend identification, prediction, and the trust game. The main survey comprised 20 tasks: four trend-identification tasks, four prediction tasks, and twelve trust-game tasks. Images from the tutorial are included in the \href{https://osf.io/x9zmr/?view_only=c3a7071971e5420ea5f8b99e5af420dd}{supplemental material}.

%% file: sec.tasks.tex
\section{Tasks}
\label{sec.tasks}

This survey aimed to evaluate three distinct facets of visualization techniques: representing trends, facilitating predictions based on historical data, and trustworthiness in decision-making. To accomplish this, we divided the survey tasks into three categories, each with different stimuli, objectives, and outcomes, which all participants completed. By assessing participants' responses in each task category, we aimed to gain insights into the effectiveness of visualization approaches for each of these critical components. A pilot study (see \href{https://osf.io/x9zmr/?view_only=c3a7071971e5420ea5f8b99e5af420dd}{supplemental material}) revealed that the standard and aggregated charts produced similar interpretation patterns, which informed our hypotheses for the main experiment.

\input{sec.tasks.trend}

\input{sec.tasks.prediction}

\input{sec.tasks.trust}

%% file: sec.tasks.trend.tex
\subsection{Trend Identification}
This task assessed participants' ability to detect and classify trends across four visualization types.  Each participant completed four trials (standard, aggregated, trellis, and spiral) in random order, each with a different dataset. For each chart, they first indicated whether a trend was present and, if so, chose one of four categories: upward, remaining the same, repeating, or downward.

\paragraph{Stimuli}
The four tasks used one standard line chart, one aggregated chart, one trellis chart, and one spiral chart. The visualizations appeared in a random order, each depicting a different dataset. Participants responded “yes” or “no” to the presence of a trend; if “yes,” they selected the trend type from the four options: upward, remaining the same, repeating, or downward. Therefore, the independent variables were \emph{visualization} and \emph{dataset}, and the dependent variables were \emph{trend visibility} and \emph{trend type}.

\paragraph{Hypothesis}
The trend identification task aimed to compare how effectively each design reveals underlying patterns. We formulated two hypotheses guided by the pilot study:

\begin{hyp}{\ref{hyp:trend}}\label{hyp:trend}
  \textit{Trends are similarly apparent in standard and aggregated charts and less apparent in trellis or spiral charts.}
\end{hyp}

\begin{hyp}{\ref{hyp:trend_correctness}}\label{hyp:trend_correctness}
  \textit{Standard and aggregated charts enable users to identify the correct trend more often than trellis or spiral charts.}
\end{hyp}

\begin{wrapfigure}{r}{0.42\linewidth}
    \centering

     \subfloat[Identified any trend\label{fig:trend-identification}]{\includegraphics[width=0.425\linewidth]{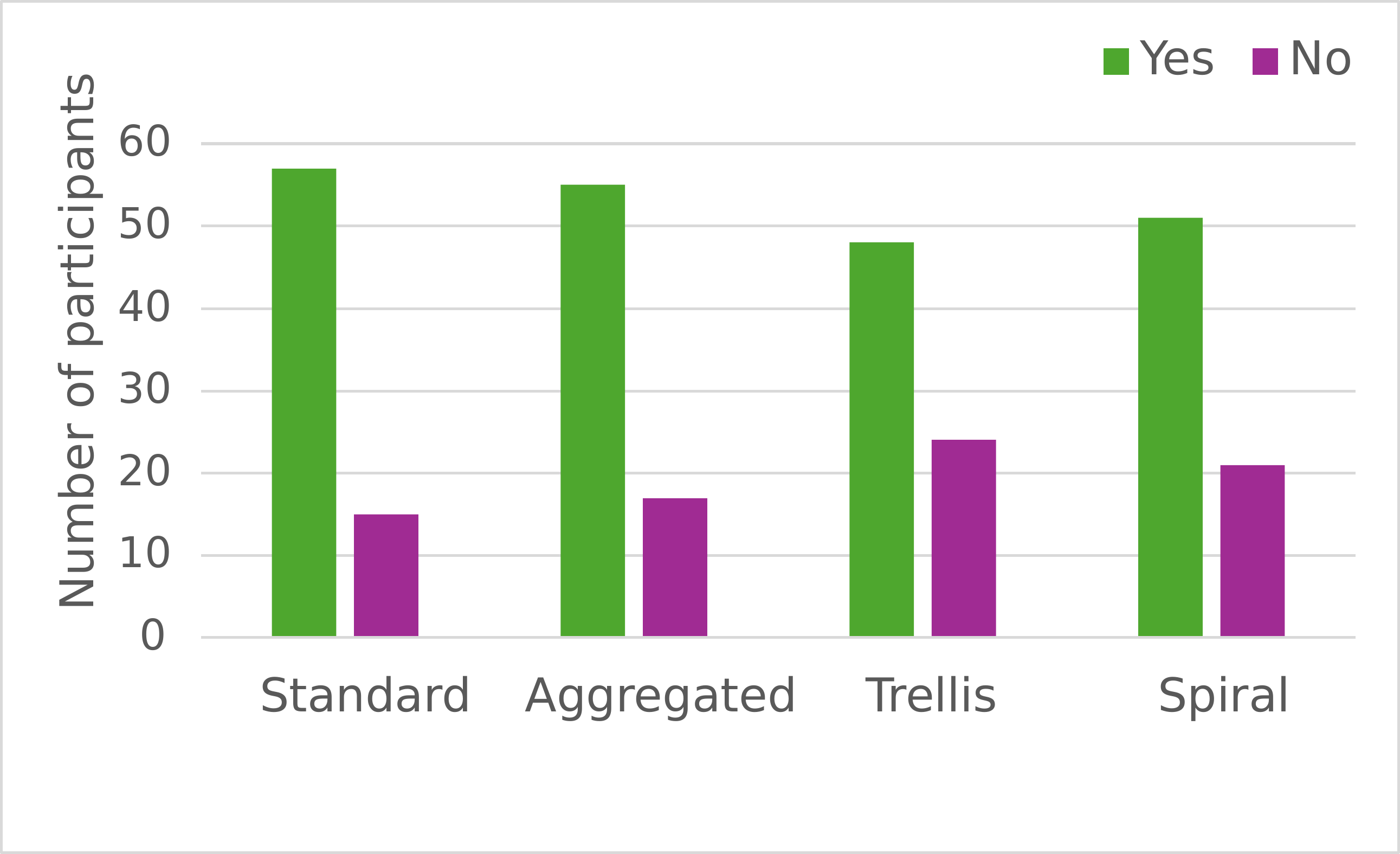}}
    \hfill
    \subfloat[Identified repeating trend\label{fig:trend-identification-periodic}]{\includegraphics[width=0.425\linewidth]{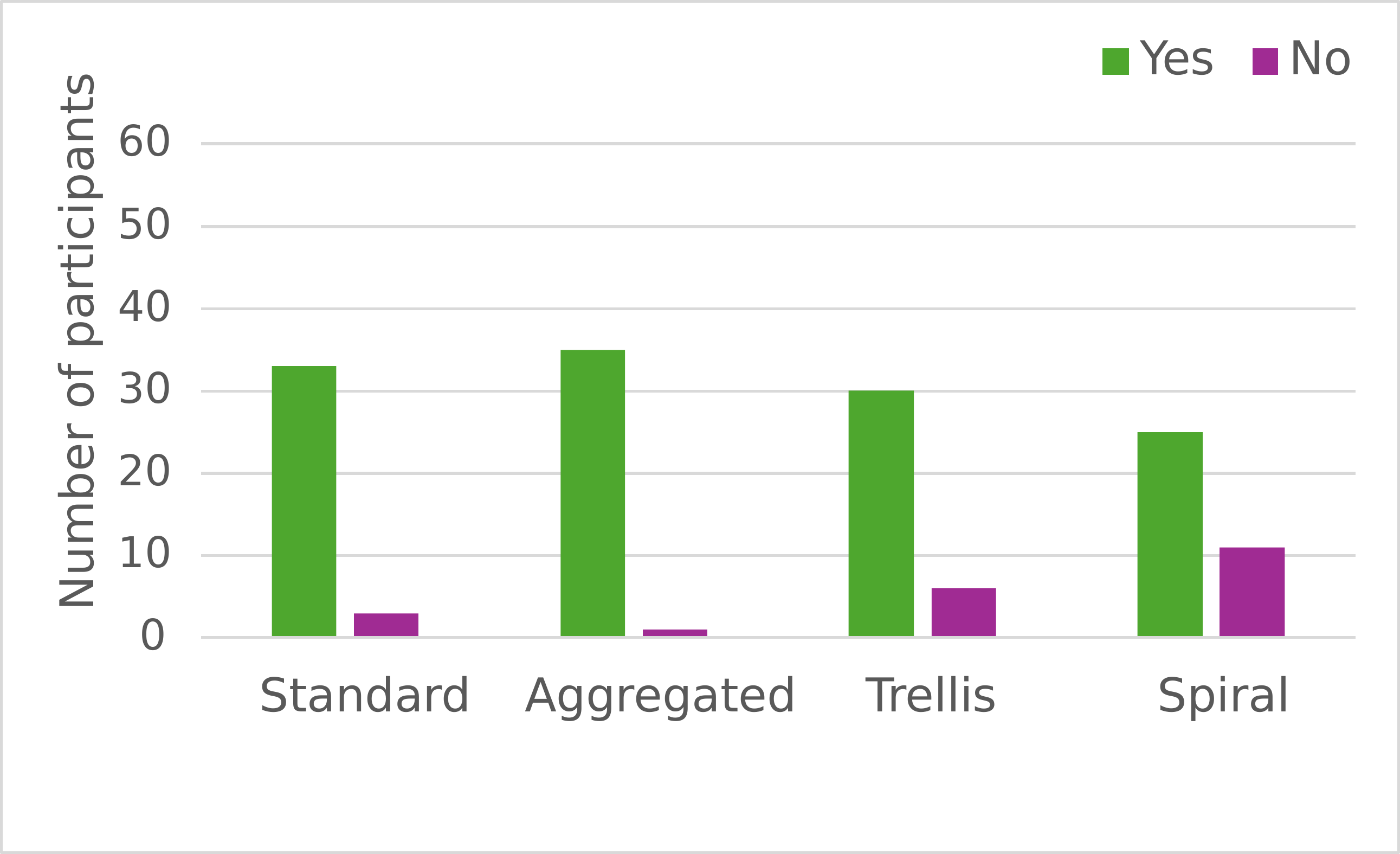}}

    \subfloat[Accuracy for any trend\label{fig:trend-correctness}]{\includegraphics[width=0.425\linewidth]{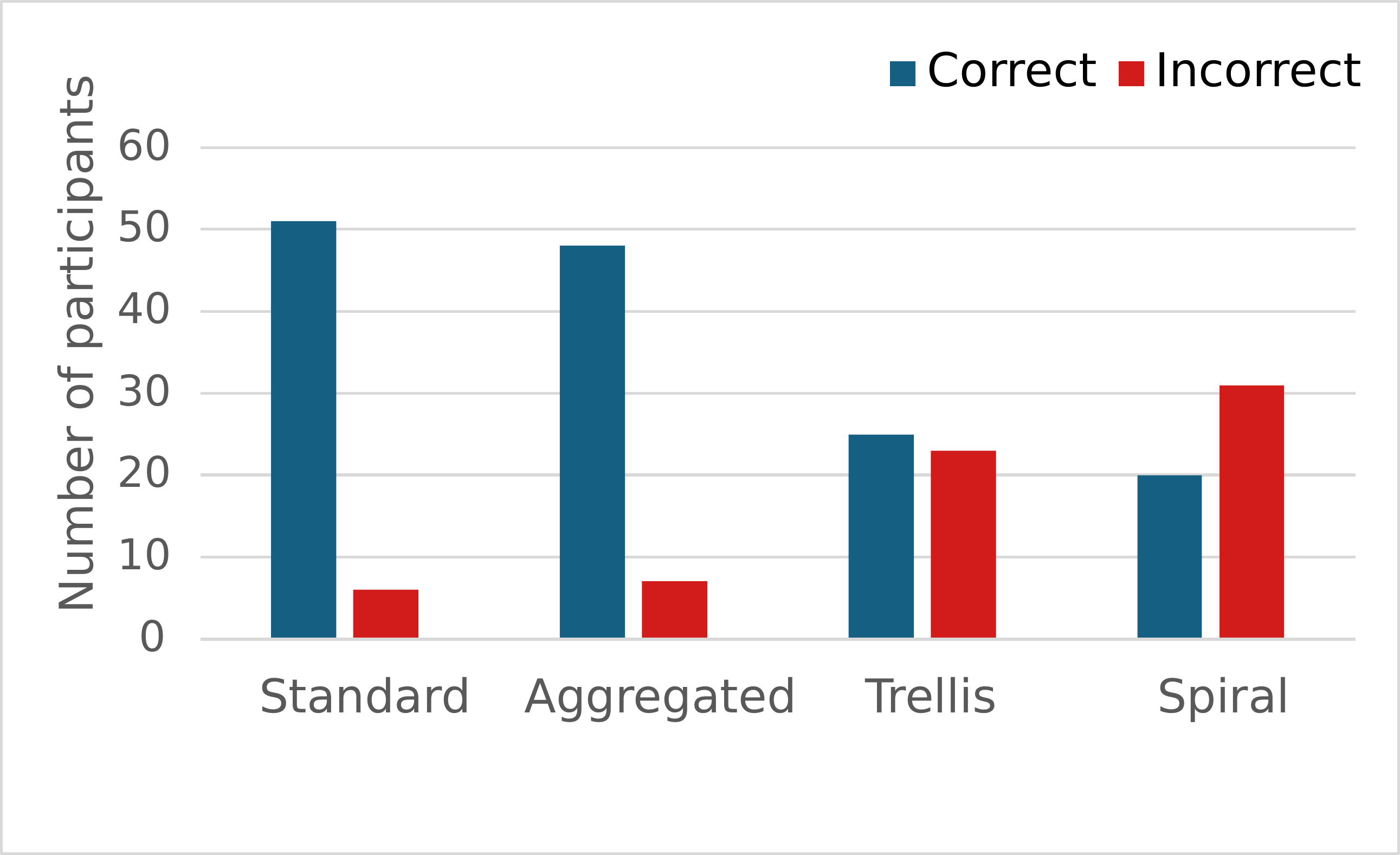}}
    \hfill
    \subfloat[Accuracy for repeating trend\label{fig:trend-correctness-periodic}]{\includegraphics[width=0.425\linewidth]{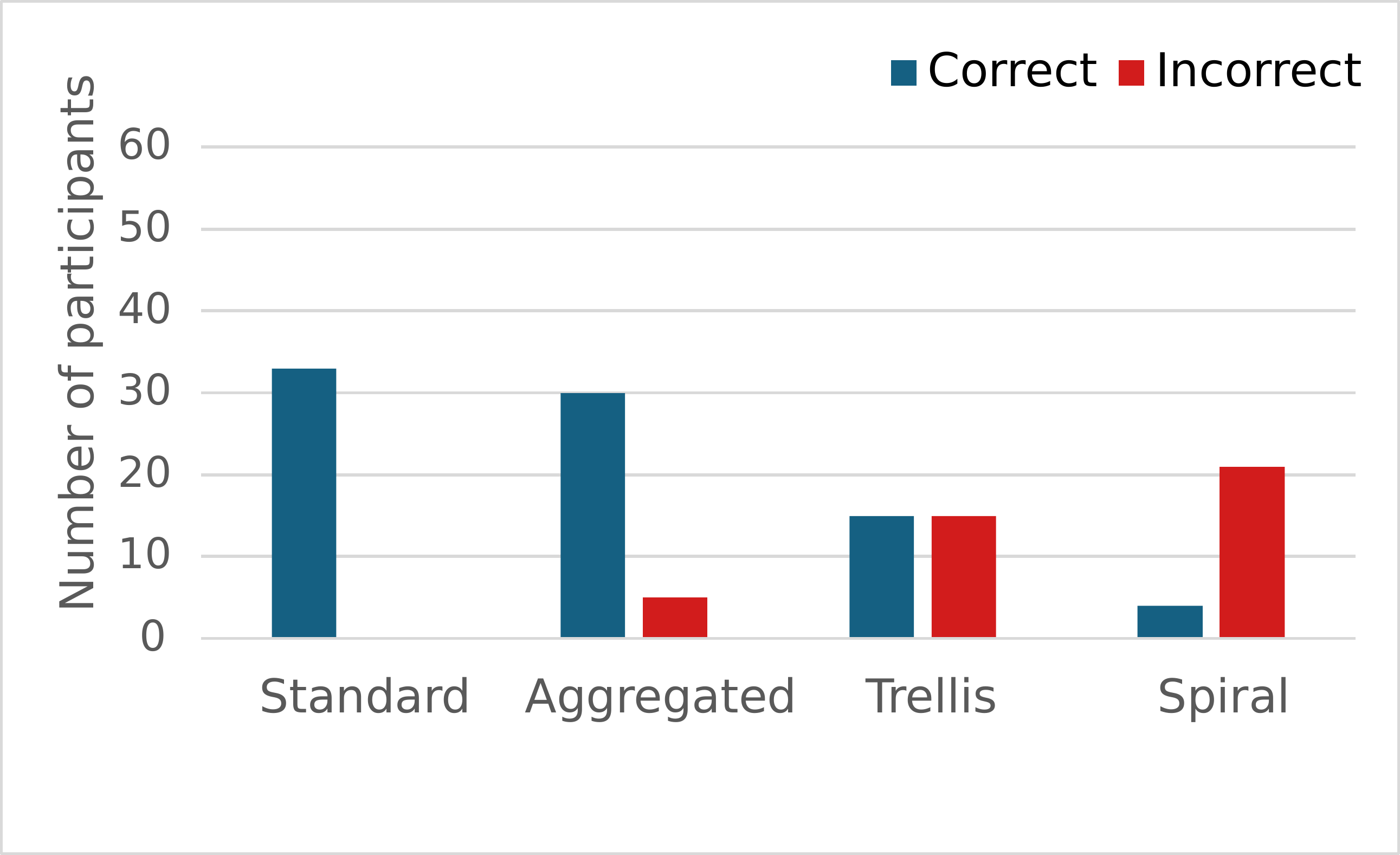}}

    \caption{Trend identification results showing whether participants detected a trend (a-b) and if that trend was correct (c-d).}
    \label{fig:trend-analysis}
\end{wrapfigure}

\paragraph{Results}
\Cref{fig:trend-analysis} summarizes the results.  A chi-square test showed no significant differences among the four designs in detecting the presence of a trend \chiSqrAPA{3}{288}{3.457}{=.326}, and pairwise comparisons against the standard chart were also nonsignificant \chiSqrAPA{1}{144}{.161}{=.688} for aggregated, \chiSqrAPA{1}{144}{2.848}{=.092}) for trellis and \chiSqrAPA{1}{144}{1.333}{=.248} for spiral charts. Thus, Hypothesis~\ref{hyp:trend} is only partially supported. However, in the periodic datasets (humidity and crime), the overall difference was significant \chiSqrAPA{3}{144}{12.655}{=.005}. The pairwise comparison showed that spiral charts differed significantly from the standard chart, whereas aggregated and trellis charts did not (see \cref{fig:trend-identification-periodic}, additional details in the \href{https://osf.io/x9zmr/?view_only=c3a7071971e5420ea5f8b99e5af420dd}{supplement}).

Trend correctness varied more noticeably (see~\cref{fig:trend-correctness}). An overall test was significant \chiSqrAPA{1}{211}{46.66}{<.001}. Aggregated and standard charts produced similar proportions of correct trend identifications \chiSqrAPA{1}{112}{0.132}{=.716}, whereas trellis and spiral charts were significantly worse  \newline \chiSqrAPA{1}{105}{18.223}{<.001} and \chiSqrAPA{1}{108}{30.187}{<.001}, 
respectively, confirming Hypothesis~\ref{hyp:trend_correctness}. These differences persisted on the periodic datasets as shown in~\Cref{fig:trend-correctness-periodic} \chiSqrAPA{3}{123}{54.844}{<.001}, with significant 
pairwise differences between standard and aggregated \chiSqrAPA{1}{68}{5.088}{=.024}, standard and trellis \chiSqrAPA{1}{63}{21.656}{<.001} and standard and spiral \chiSqrAPA{1}{58}{43.453}{<.001} charts.

\paragraph{Discussion}
Participants reported seeing trends equally often in all chart types, but their ability to identify the \emph{correct} trend dropped noticeably for trellis and spiral charts. Aggregated charts matched standard charts in both detection and correctness, demonstrating that smoothing dense series does not compromise high‑level understanding. Trellis and spiral charts, however, made it harder to identify correct trends, highlighting that aggregation is the safer alternative for conveying trends clearly.

%% file: sec.tasks.prediction.tex
\subsection{Prediction Tendency}
\label{sec:task:prediction}

This task assessed how well participants could forecast the next data point. Each trial paired one of the four chart types (standard, aggregated, trellis, or spiral) with a different dataset in random order. After viewing a chart, participants predicted whether the series would go upward, remain the same, or go downward one month later and rated their confidence on a five-point scale. Thus, the independent variables were \textit{visualization} and \textit{dataset}; the dependent variables were \textit{prediction} and \textit{confidence level}.

\paragraph{Stimuli}
Each prediction task showed a standard, aggregated, trellis, or spiral chart exactly once with a brief description of its content. Chart orders were randomized, with each chart showing a different dataset. After examining the chart, participants selected “it will go upwards,” “it will remain the same” or “it will go downwards,” and rated their confidence (“very sure,” “sure,” “neither sure nor unsure,” “unsure”, or “very unsure”).

\paragraph{Hypothesis}
For prediction, we examined how the three alternative charts compare to a standard line chart in supporting consistent predictions. Based on our observations in the pilot study, we formulated the following hypothesis:

\begin{hyp}{\ref{hyp:prediction_accuracy}}\label{hyp:prediction_accuracy}
\textit{Participants will make similar predictions with standard and aggregated charts, while predictions on trellis and spiral charts will differ.}
\end{hyp}

\begin{hyp}{\ref{hyp:prediction_confidence}}\label{hyp:prediction_confidence}
\textit{Participants will have similar confidence in their predictions when using standard and aggregated line charts, while trellis and spiral charts will be lower.}
\end{hyp}

\begin{wrapfigure}{r}{0.31\linewidth}
  \centering
    \subfloat[Prediction per charts\label{fig:prediction-consistency}]{\includegraphics[width=0.975\linewidth]{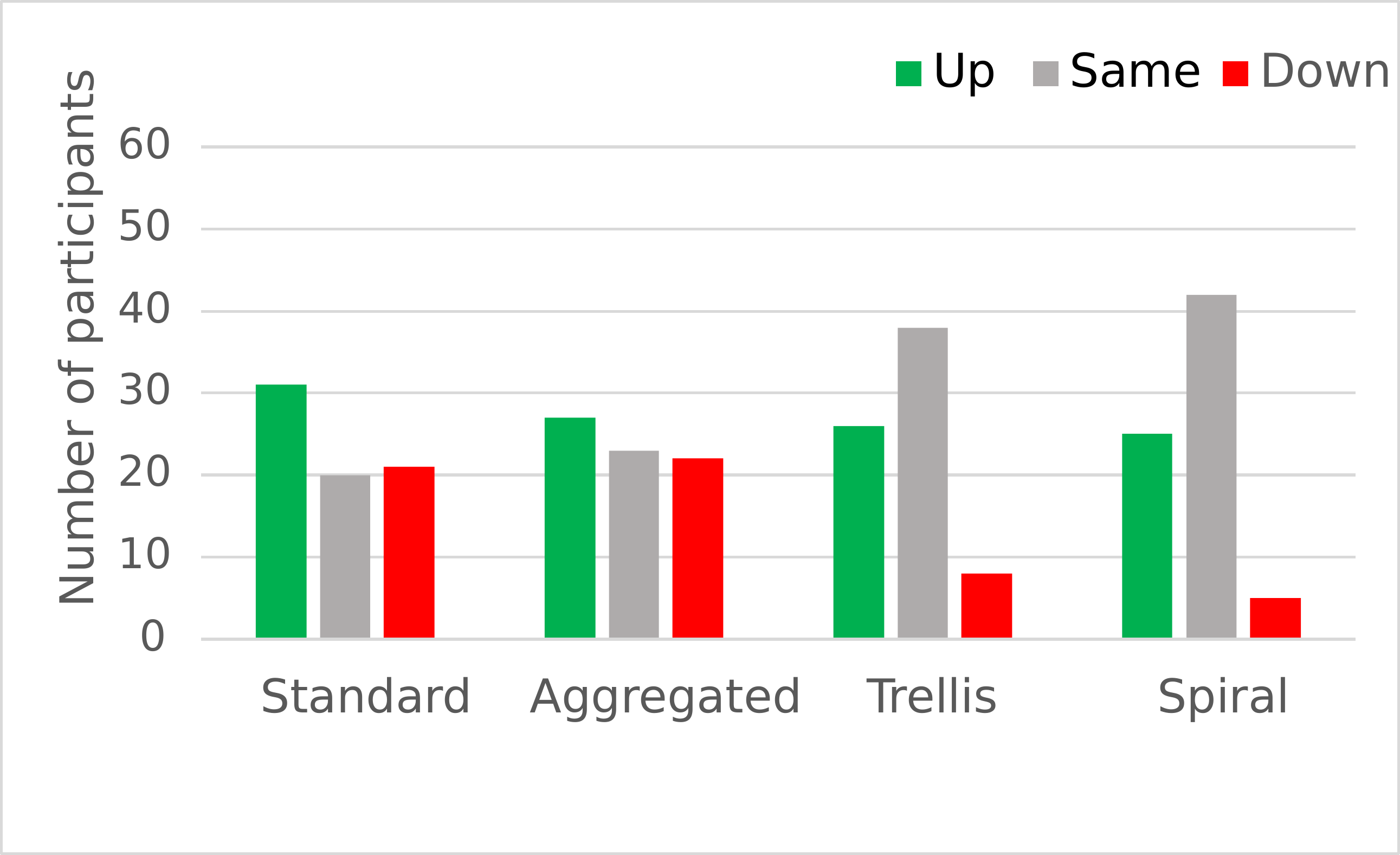}}
    
    \subfloat[Confidence of prediction\label{fig:prediction-confidence}]{\includegraphics[width=0.975\linewidth]{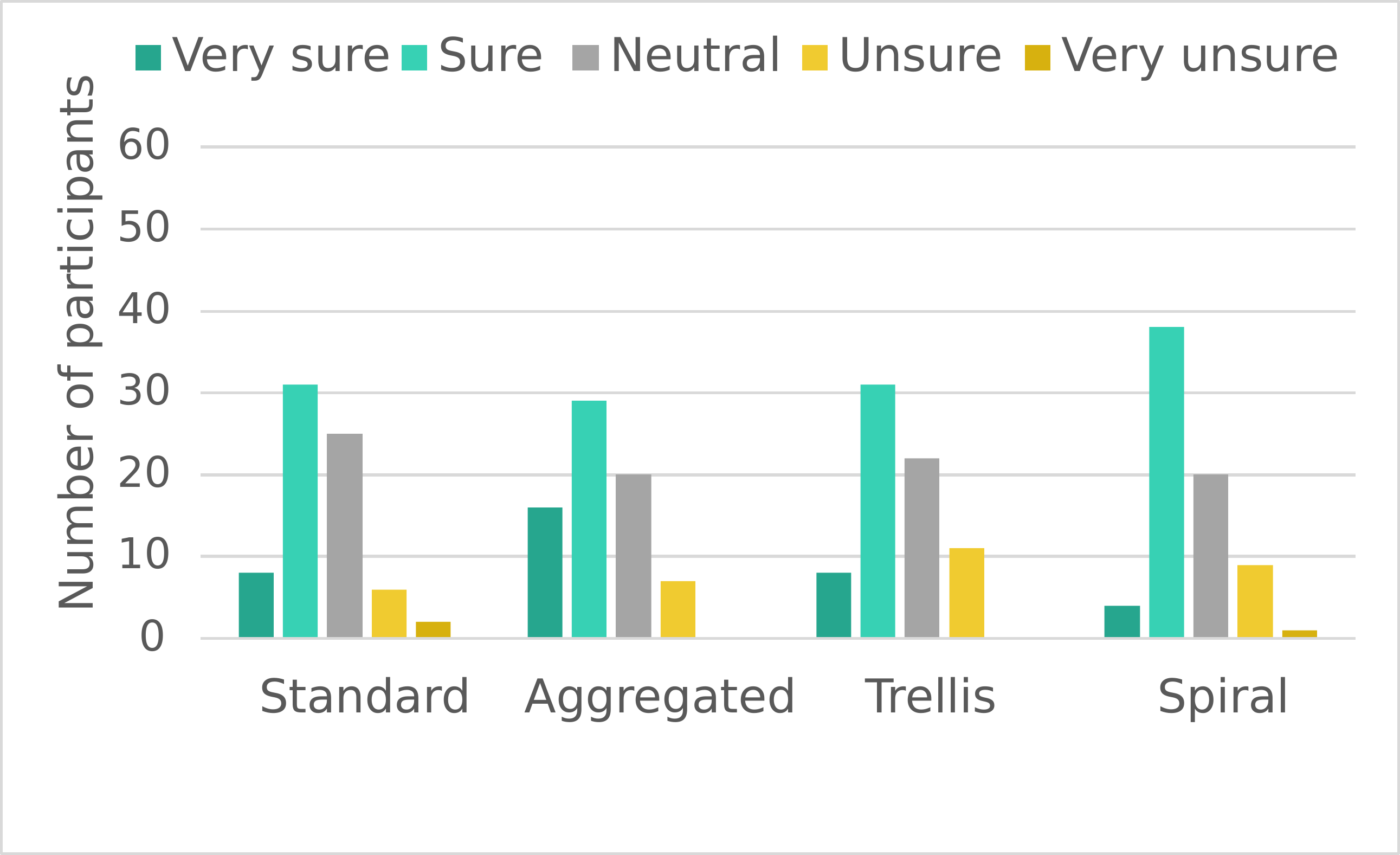}}
  \caption{Results from the prediction tasks.}
  \label{fig:prediction-analysis}
\end{wrapfigure}

\paragraph{Results}
A chi-square test revealed a significant difference among the four visualization approaches \chiSqrAPA{6}{288}{28.727}{<.001}. Predictions from standard and aggregated charts did not differ \chiSqrAPA{2}{144}{0.508}{=.776}, whereas predictions differed significantly for trellis \chiSqrAPA{2}{144}{11.852}{=.003} and spiral \chiSqrAPA{2}{144}{18.296}{<.001} charts compared to the standard. These findings confirm Hypothesis~\ref{hyp:prediction_accuracy}.

We also examined participants’ confidence in their predictions. Responses were coded from 1 (“very sure”) to 5 (“very unsure”), with 3 representing a neutral stance. Mean scores across all designs were between “sure” and “neither sure nor unsure”, indicating moderate confidence. Analysis using t-test revealed no significant difference in confidence between standard and aggregated charts \tTestAPA{142}{1.56}{=.122} (\MeanSdAPA{2.49}{.904} vs.\ \MeanSdAPA{2.25}{0.915}), nor between standard and trellis \tTestAPA{142}{-0.093}{=.926} \MeanSdAPA{2.50}{.888} or standard and spiral \tTestAPA{142}{-0.191}{=.849} \MeanSdAPA{2.51}{.839}. Thus, Hypothesis~\ref{hyp:prediction_confidence} is rejected. \Cref{fig:prediction-analysis} shows prediction outcomes and confidence ratings for each chart.

\paragraph{Discussion}
Aggregated charts preserved prediction patterns relative to the standard chart, whereas trellis and spiral charts led to significantly different forecasts. Across all chart types, confidence levels did not significantly vary. On average, participants were mostly sure about their predictions. This is important (\emph{and potentially problematic}), considering that predictions appear biased for trellis and spiral charts.

%% file: sec.tasks.trust.tex
\subsection{Decision-Making and Trust}
\label{sec:task:trust}

Trust is essential for visualizations to support informed decision-making. We adapted the trust-game approach by Elhamdadi et al.~\cite{elhamdadi2022using} for our experiment. In each round, participants saw two charts (a standard chart vs. an aggregated, trellis, or spiral) and decided how to invest \$100 between two companies. Higher allocation toward one indicated higher trust in that chart.

\begin{wrapfigure}{r}{0.49\linewidth}

        \centering
        \subfloat[Both charts show an upward trend; the aggregated chart (bottom) rises slightly faster.]{\begin{minipage}[b]{0.5\linewidth}
            \centering
            Company A
            \includegraphics[width=2.25cm]{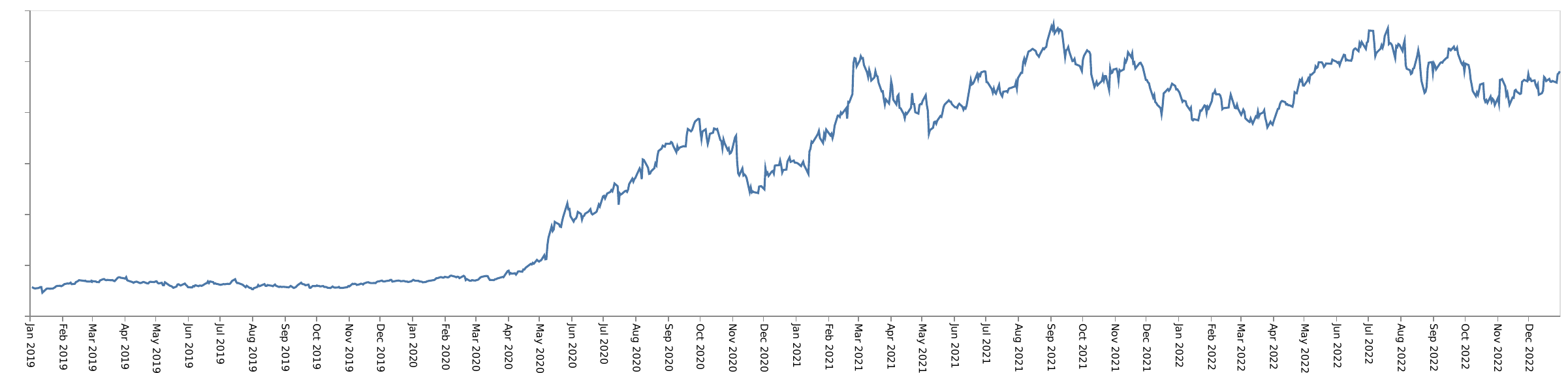}

            Company B
            \includegraphics[width=2.25cm]{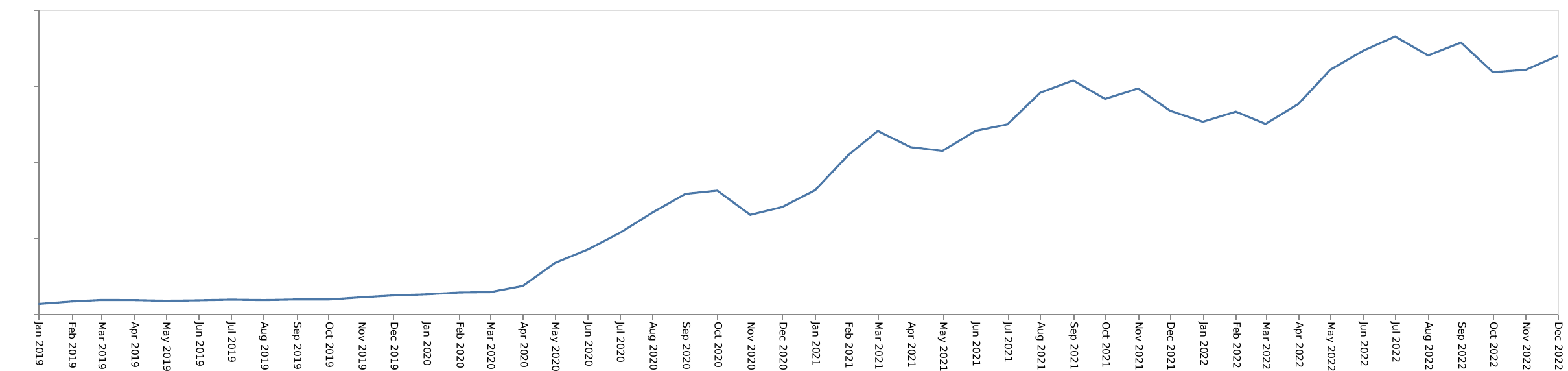}
        \end{minipage}}
        \hfill
        \subfloat[Charts show opposite trends: standard (top) down, aggregated (bottom) up.]{\begin{minipage}[b]{0.44\linewidth}
            \centering        
            Company A
            \includegraphics[width=2.25cm]{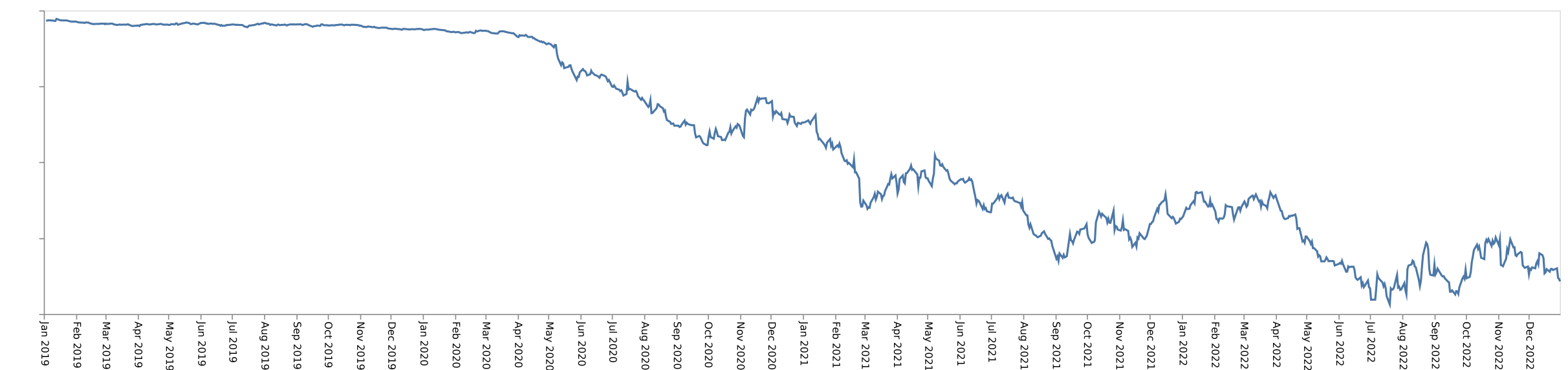}
            Company B
            \includegraphics[width=2.25cm]{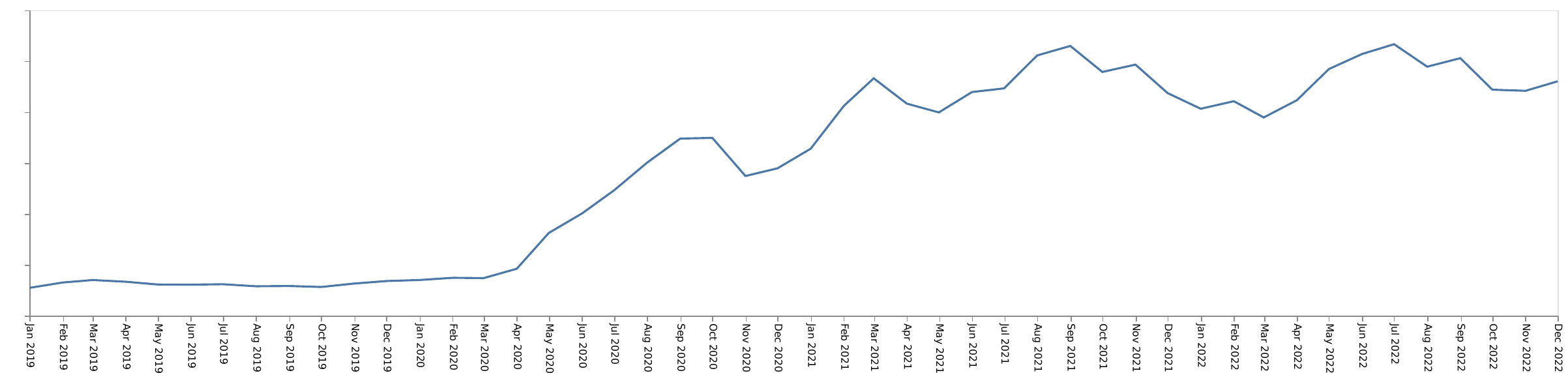}
        \end{minipage}}

        \subfloat[Both standard (top) and trellis (bottom) show the same downward trend.]{
        \begin{minipage}[b]{0.475\linewidth}
            \centering
            Company A
            \includegraphics[width=2.25cm]{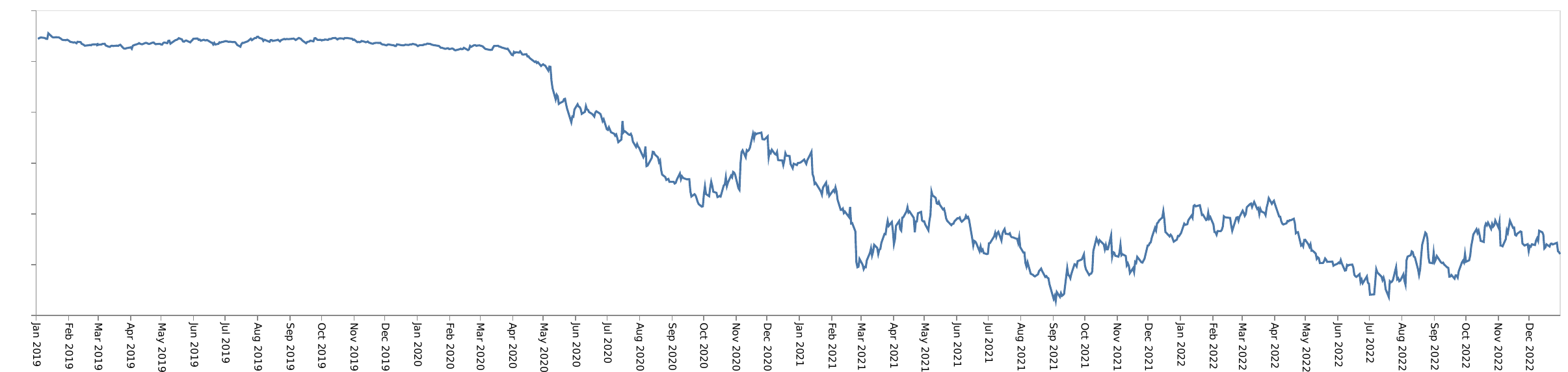}
            Company B
            \includegraphics[trim=20pt 0 0 0, clip, height=1.5cm]{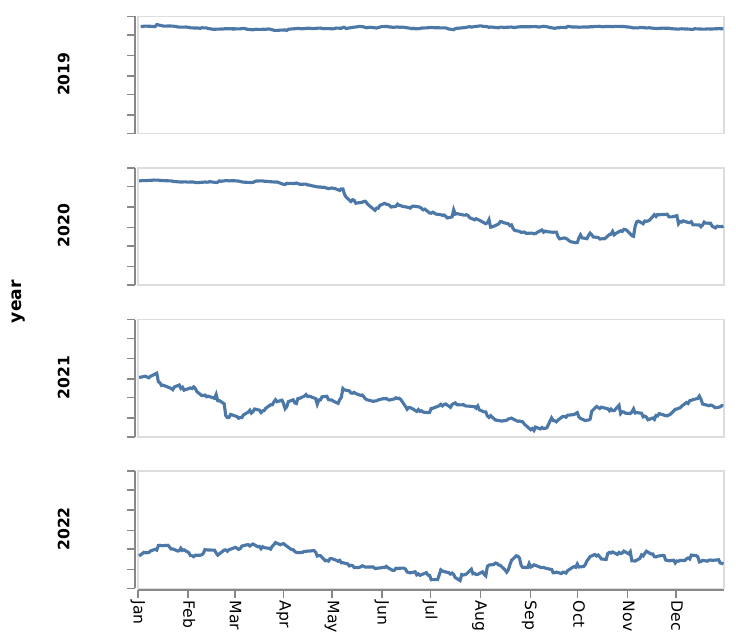}
        \end{minipage}
        }
        \hfill
        \subfloat[Opposite trends: standard (top) up, spiral (bottom) down.]{
            \begin{minipage}[b]{0.375\linewidth}
                \centering
                Company A
                \includegraphics[width=2.25cm]{figs/results/trust/daily_up_trust_game.pdf}
                Company B
                \includegraphics[height=1.5cm]{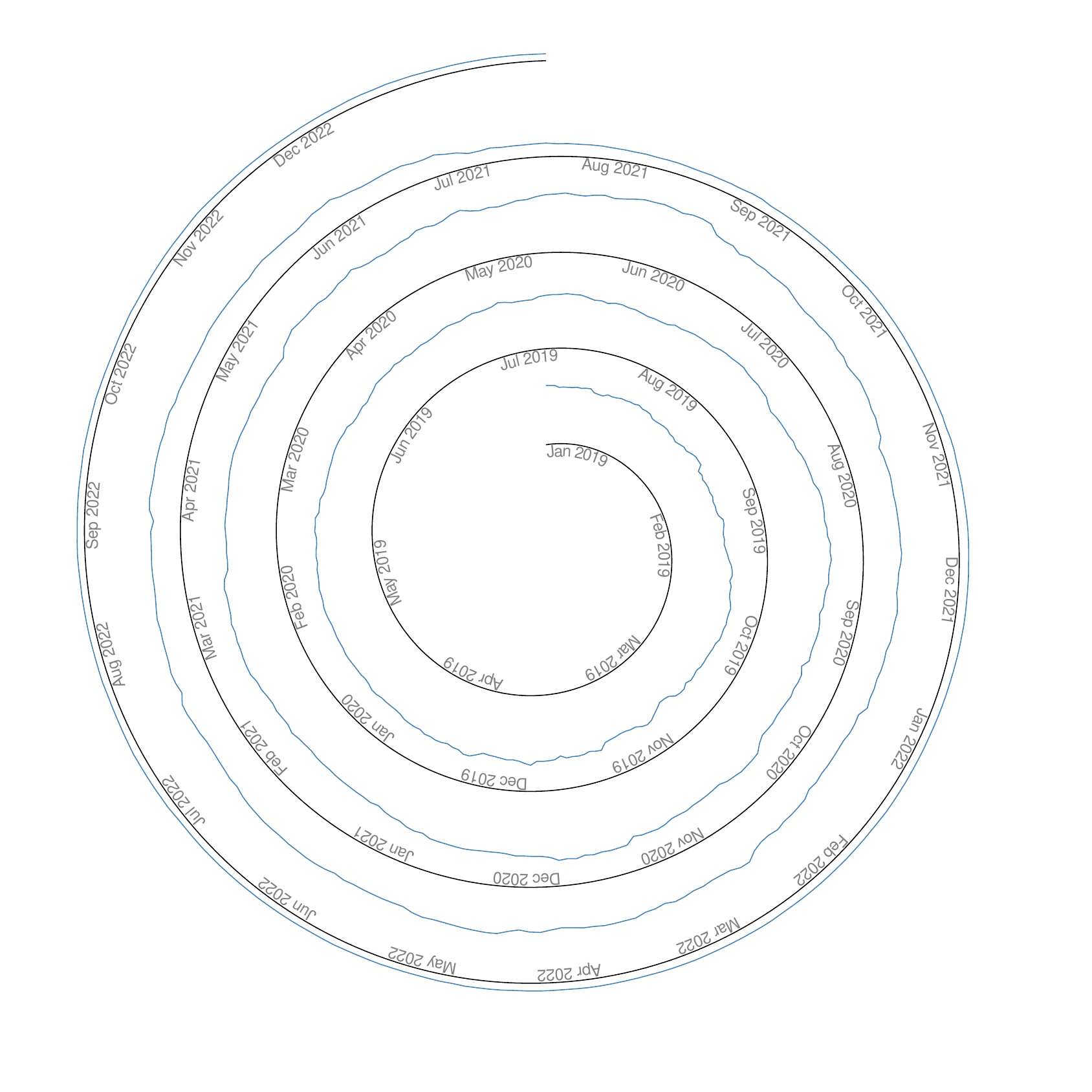}
            \end{minipage}
        }

    \caption{Examples of chart pairs used in the trust game. Each participant played 12 rounds, each with one pair.}
    \label{fig:trust-game-example}
\end{wrapfigure}

\paragraph{Stimuli}
We used Tesla stock data from Kaggle, manipulated to exhibit specific trends. The original dataset was used for upward movement, its values negated for downward movement, and a growth factor was applied to 
create faster changes. The objective was to compare aggregated, trellis, and spiral charts against a standard line chart that plotted all data points. To this end, we designed a trust game in which each participant played 12 rounds. Each round presented a pair of visualizations—a standard chart 
and one alternative design as illustrated in~\Cref{fig:trust-game-example}. We created four trend types and six groups (see~\Cref{table:trust_game_group}), which were combined into eight different configurations (see~\Cref{table:trust_game_group_combination}). We targeted 80 participants overall, assigning 10 to each of the eight groups. Group 
order was randomized to avoid ordering effects. The independent variables were \emph{visualization type} and \emph{trend type}; the dependent variable was the participant’s \emph{investment fund allocation}.

\begin{table*}[!b]
    \begin{minipage}[t]{0.6\textwidth}
        \caption{Stimuli groups used in the trust game.}
        \label{table:trust_game_group}
        \resizebox{\linewidth}{!}{%
        \begin{tabular}{cc|c|c|c|c|c}
        Standard Chart               & Group  & \multirow{2}{*}{Visualization}       & \multicolumn{2}{c|}{Trends same direction}    & \multicolumn{2}{c}{Trends opposite direction} \\
        compared to                       & ID     &                                      &   Type 1                     & Type 2         & Type 3               & Type 4                 \\
        \hline\hline
                                          & \multirow{2}{*}{$A_1$} & Standard Chart & \multirow{2}{*}{$\uparrow$ Same Rate}   & $\uparrow$ Higher Rate & $\downarrow$ Same Rate       & $\downarrow$ Higher Rate       \\
    Aggregated                            & & Aggregated Chart    &                                 & $\uparrow$ Lower Rate  & $\uparrow$ Same Rate         & $\uparrow$ Lower Rate          \\
                                          \cline{2-7}
    Chart                           & \multirow{2}{*}{$A_2$} & Standard Chart & \multirow{2}{*}{$\downarrow$ Same Rate} & $\uparrow$ Lower Rate  & $\uparrow$ Same Rate         & $\downarrow$ Higher Rate       \\
                                          &                      & Aggregated Chart    &                                 & $\uparrow$ Higher Rate & $\downarrow$ Same Rate       & $\uparrow$ Lower Rate          \\
        \hline
                                & \multirow{2}{*}{$T_1$} & Standard Chart & \multirow{2}{*}{$\uparrow$ Same Rate}   & $\uparrow$ Higher Rate & $\downarrow$ Same Rate       & $\downarrow$ Higher Rate       \\
        Trellis                    &                      & Trellis Chart       &                                 & $\uparrow$ Lower Rate  & $\uparrow$ Same Rate         & $\uparrow$ Lower Rate          \\
                                          \cline{2-7}
        Chart                        & \multirow{2}{*}{$T_2$} & Standard Chart & \multirow{2}{*}{$\downarrow$ Same Rate} & $\uparrow$ Lower Rate  & $\uparrow$ Same Rate         & $\downarrow$ Higher Rate       \\
                                          &                      & Trellis Chart       &                                 & $\uparrow$ Higher Rate & $\downarrow$ Same Rate       & $\uparrow$ Lower Rate          \\
        \hline
                          & \multirow{2}{*}{$S_1$} & Standard Chart & \multirow{2}{*}{$\uparrow$ Same Rate}   & $\uparrow$ Higher Rate & $\downarrow$ Same Rate       & $\downarrow$ Higher Rate       \\
        Spiral                       &                      & Spiral Chart        &                                 & $\uparrow$ Lower Rate  & $\uparrow$ Same Rate         & $\uparrow$ Lower Rate          \\
                                          \cline{2-7}
        Chart                                  & \multirow{2}{*}{$S_2$} & Standard Chart & \multirow{2}{*}{$\downarrow$ Same Rate} & $\uparrow$ Lower Rate  & $\uparrow$ Same Rate         & $\downarrow$ Higher Rate       \\
                                          &                      & Spiral Chart        &                                 & $\uparrow$ Higher Rate & $\downarrow$ Same Rate       & $\uparrow$ Lower Rate          \\
        \end{tabular}}    
    \end{minipage}
    \hfill
    \begin{minipage}[t]{0.35\textwidth}
        \caption{Combinations used in trust game.}
        \label{table:trust_game_group_combination}
        \resizebox{0.90\linewidth}{!}{%
        \begin{tabular}{c|ccc}
        Combination & Aggregated & Trellis & Spiral \\
        \hline\hline
        1 & $A_1$ & $T_1$ & $S_1$ \\
        2 & $A_1$ & $T_1$ & $S_2$ \\
        3 & $A_1$ & $T_2$ & $S_1$ \\
        4 & $A_1$ & $T_2$ & $S_2$ \\
        5 & $A_2$ & $T_1$ & $S_1$ \\
        6 & $A_2$ & $T_1$ & $S_2$ \\
        7 & $A_2$ & $T_2$ & $S_1$ \\
        8 & $A_2$ & $T_2$ & $S_2$ \\
        \end{tabular}}    
    \end{minipage}
\end{table*}

\paragraph{Procedure}
Participants received a scenario: invest \$100 in company A (standard) and company B (alternative) stocks with the goal of selling next month. Allocation options were - ``Spend only in company A'', ``Spend more in company A'', ``Spend an equal amount in company A and B'', ``Spend more in company B'', and ``Spend only in company B". In each round, participants saw a standard line chart paired with an alternative visualization. The two charts either showed the same trend (both rising or falling at identical or different slopes) or opposite trends (one increased while the other decreased, with either slope potentially steeper). 
Placing more money in a chart type is interpreted as placing greater trust in that visualization.
Therefore, if users have equal trust in standard line charts and the alternative charts from \Cref{table:trust_game_group}, for Type 1, we expected them to spend an equal amount of money on both charts. For Type 2, we expect them to spend more money on the visualization showing an upward trend at a higher rate. For Type 3 and Type 4, we expect investment only in the upward trend visualization.

\begin{figure}[!t]
    \subfloat[Type 1 - both charts $\uparrow$ same rate\label{fig:both_up_same}]{\includegraphics[width=0.19\linewidth]{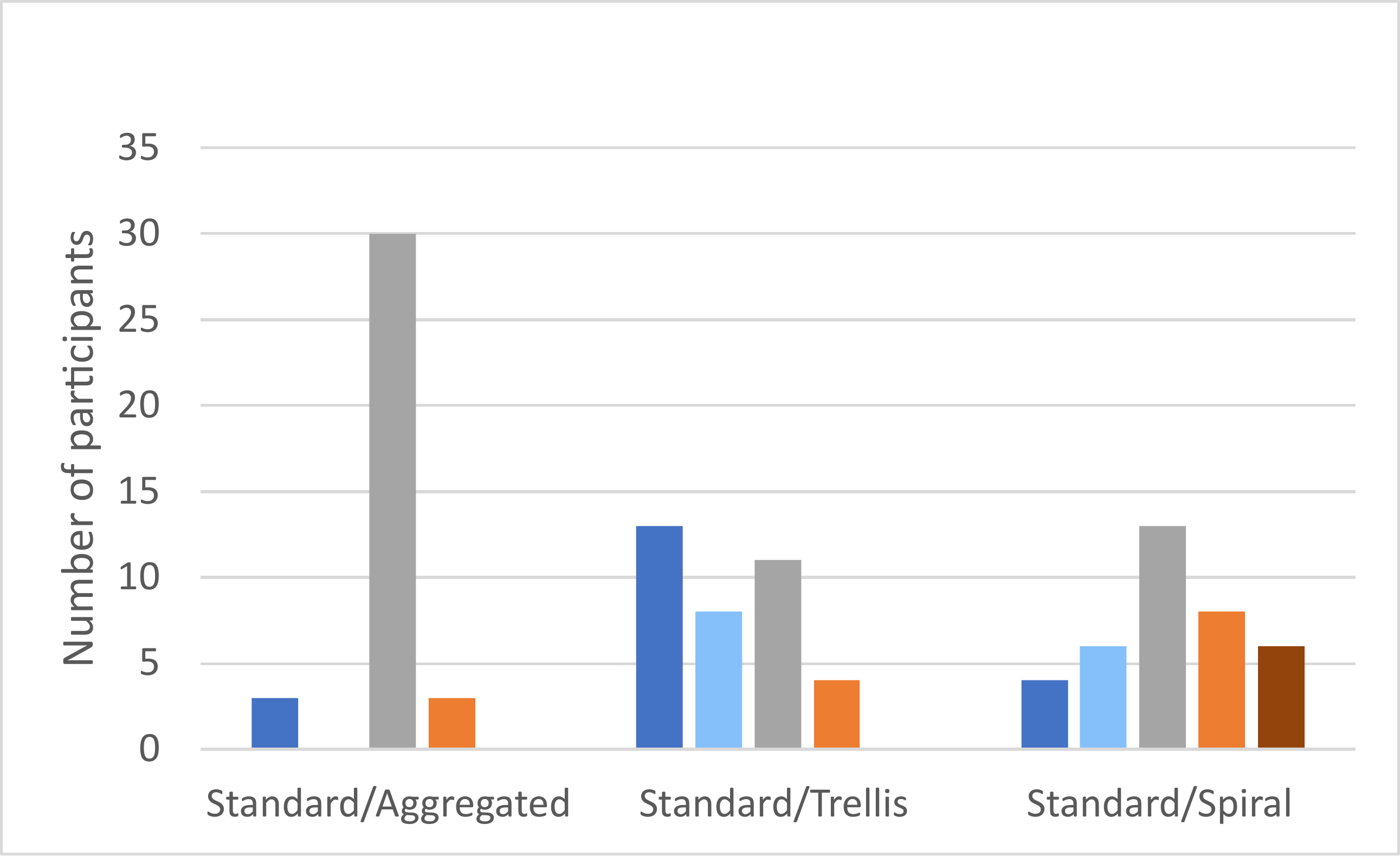}}
    \hspace{5pt}
    \subfloat[Type 1 - both charts $\downarrow$ same rate\label{fig:both_down_same}]{\includegraphics[width=0.19\linewidth]{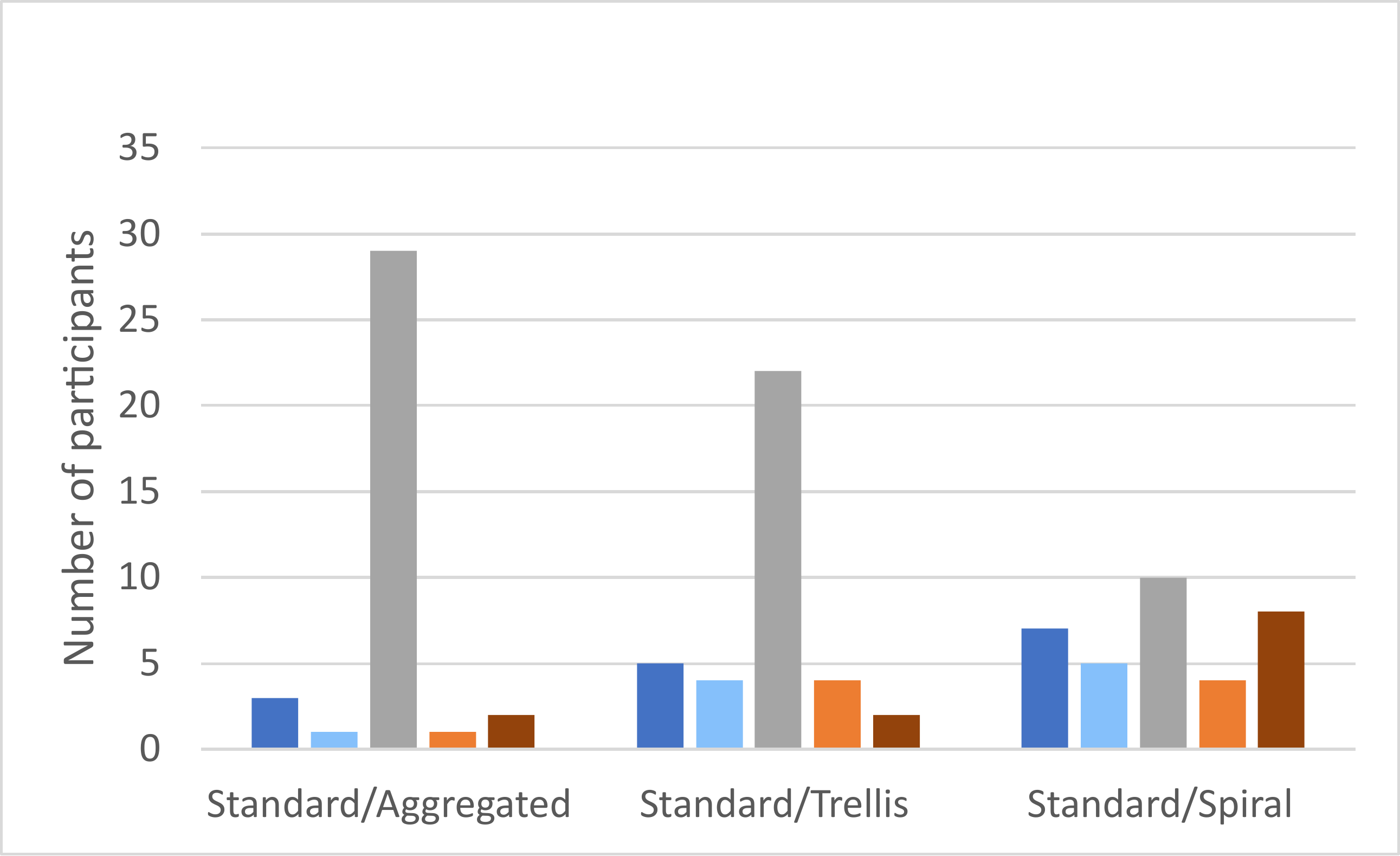}}
     \hspace{5pt}
    \subfloat[Type 2 - std.\ chart $\uparrow$ higher rate\label{fig:std_up_higher}]
    {\includegraphics[width=0.19\linewidth]{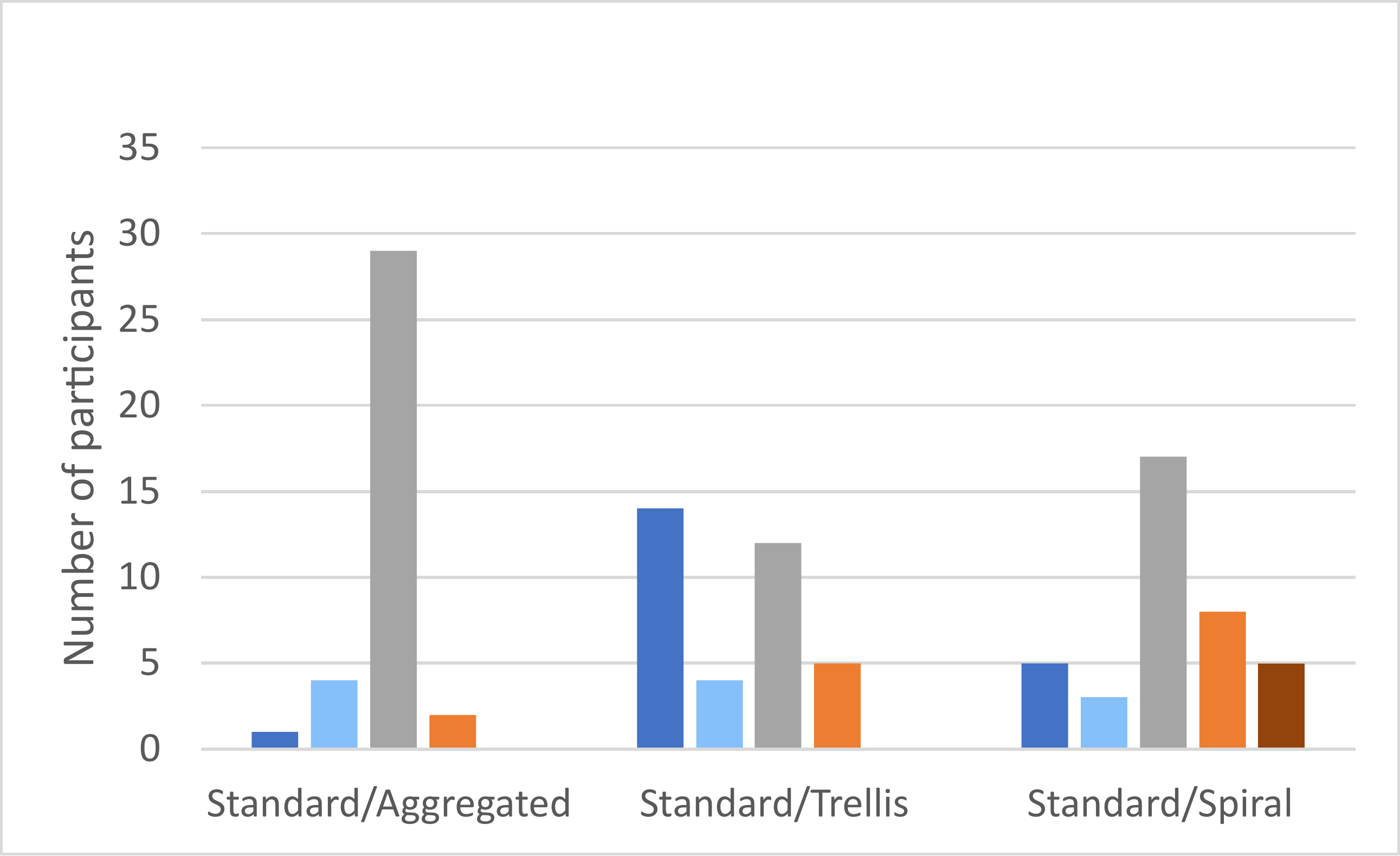}}
    \hspace{5pt}
    \subfloat[Type 2 - alt.\ chart $\uparrow$ higher rate\label{fig:other_up_higher}]{\includegraphics[width=0.19\linewidth]{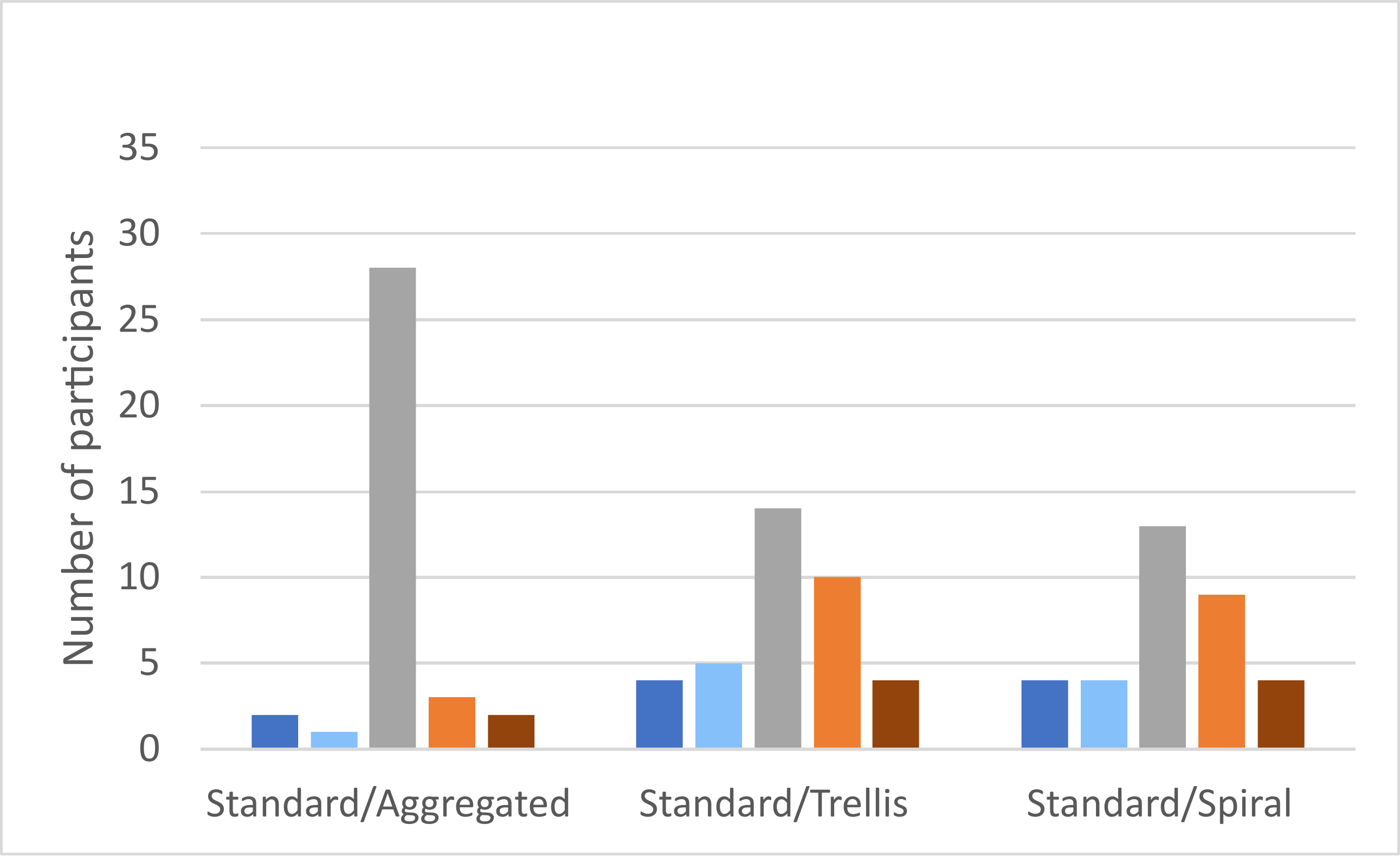}}

    \subfloat[Type 3 - std.\ chart $\downarrow$ / alt.\ chart $\uparrow$ same rate\label{fig:std_down_same}]{\includegraphics[width=0.19\linewidth]{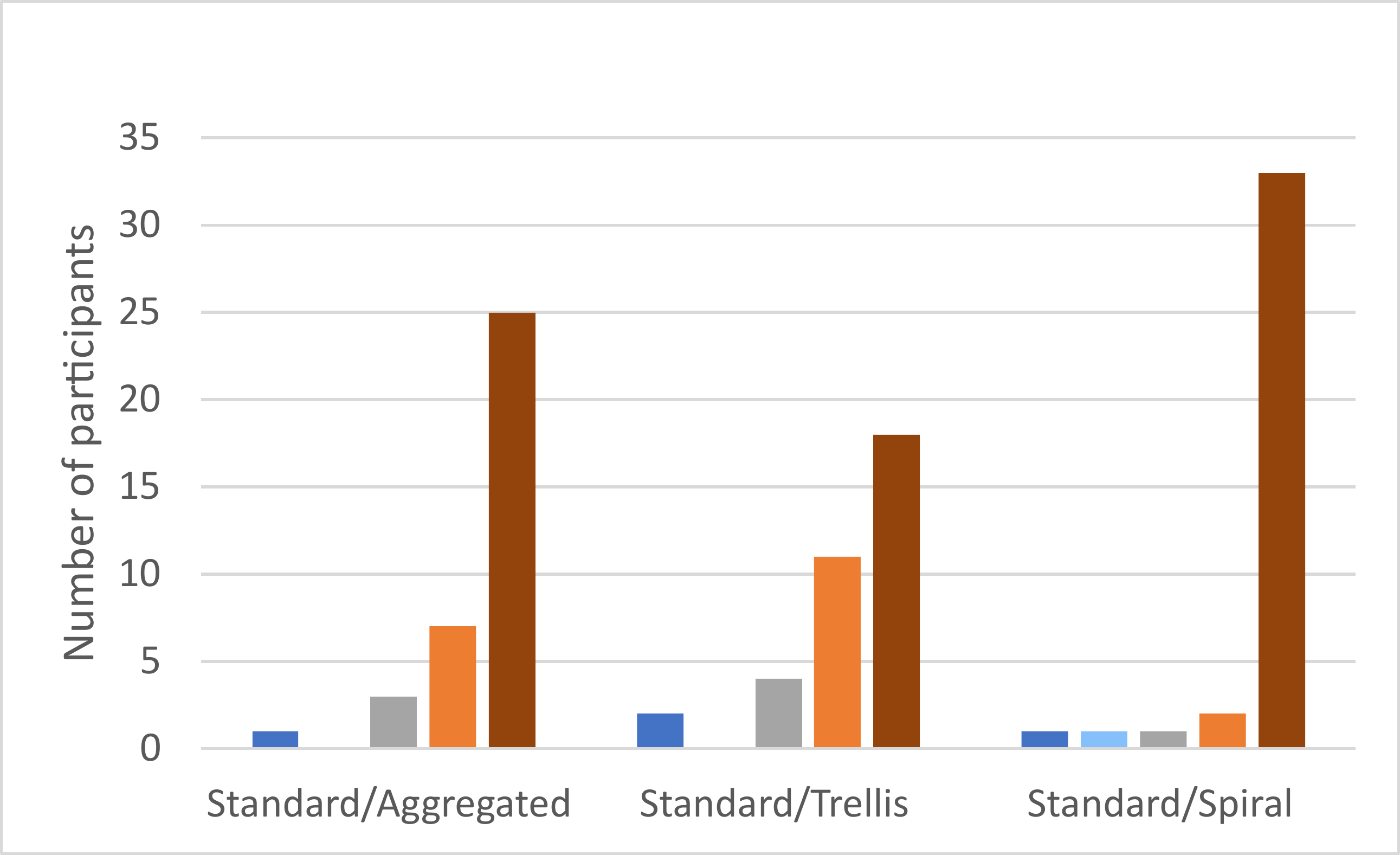}}
    \hspace{5pt}
    \subfloat[Type 3 - std.\ chart $\uparrow$ / alt.\ chart $\downarrow$ same rate\label{fig:other_down_same}]{\includegraphics[width=0.19\linewidth]{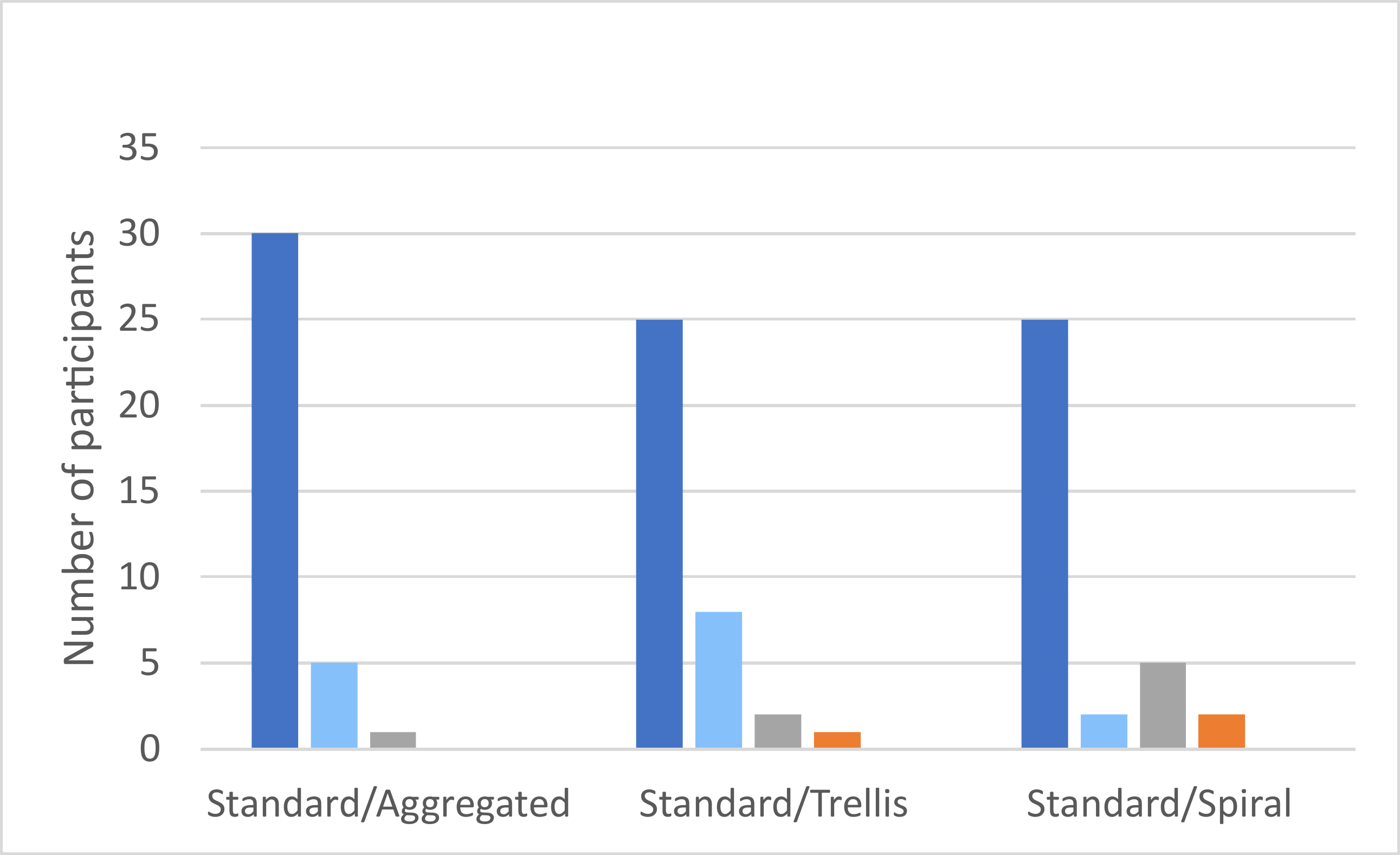}}
    \hspace{5pt}
    \subfloat[Type 4 - alt.\ chart $\uparrow$ / std.\ chart $\downarrow$ higher rate\label{fig:std_down_higher}]
    {\includegraphics[width=0.19\linewidth]{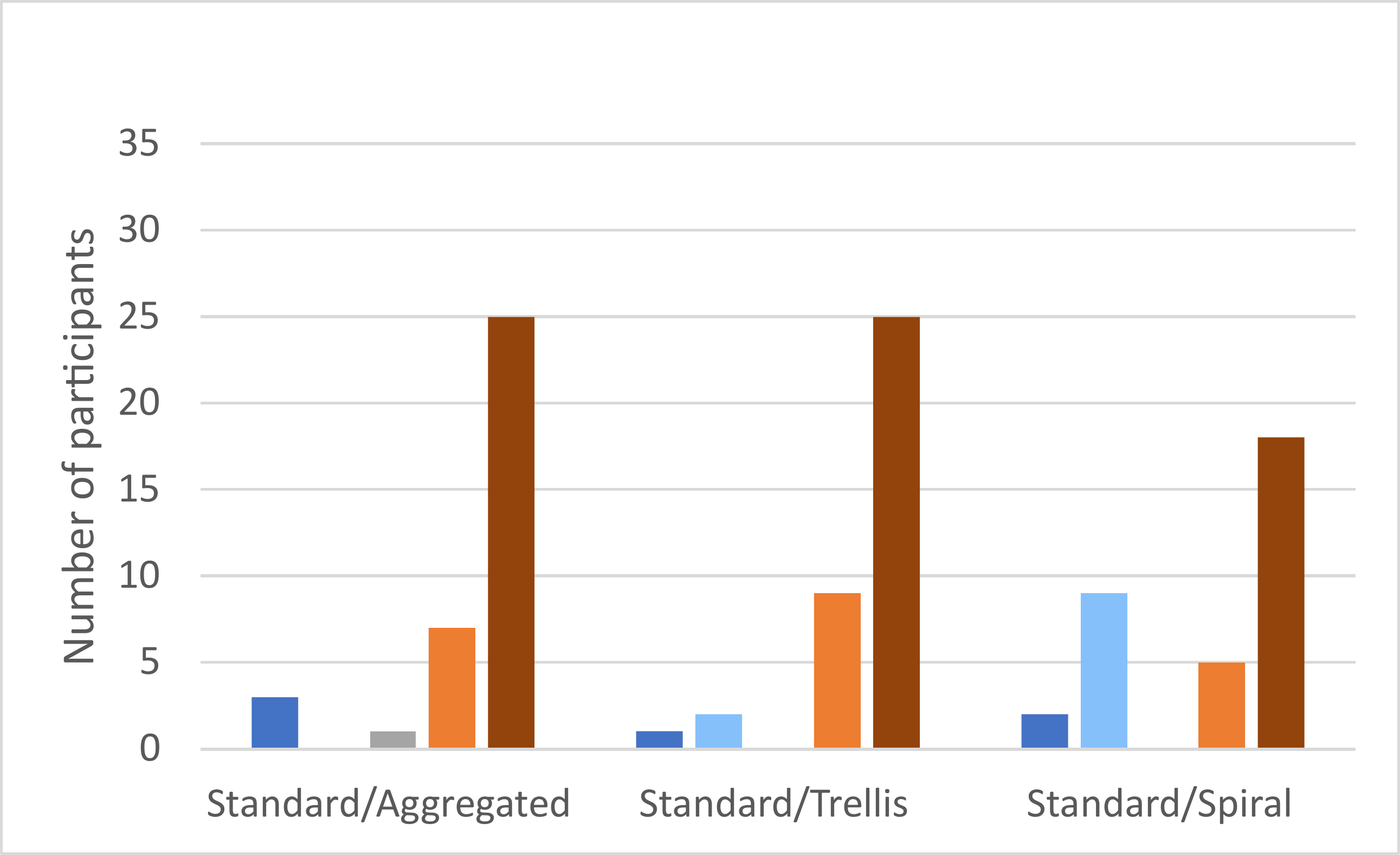}}
    \hspace{5pt}
    \subfloat[Type 4 - std.\ chart $\uparrow$ / alt.\ chart $\downarrow$ higher rate\label{fig:other_down_higher}]{\includegraphics[width=0.19\linewidth]{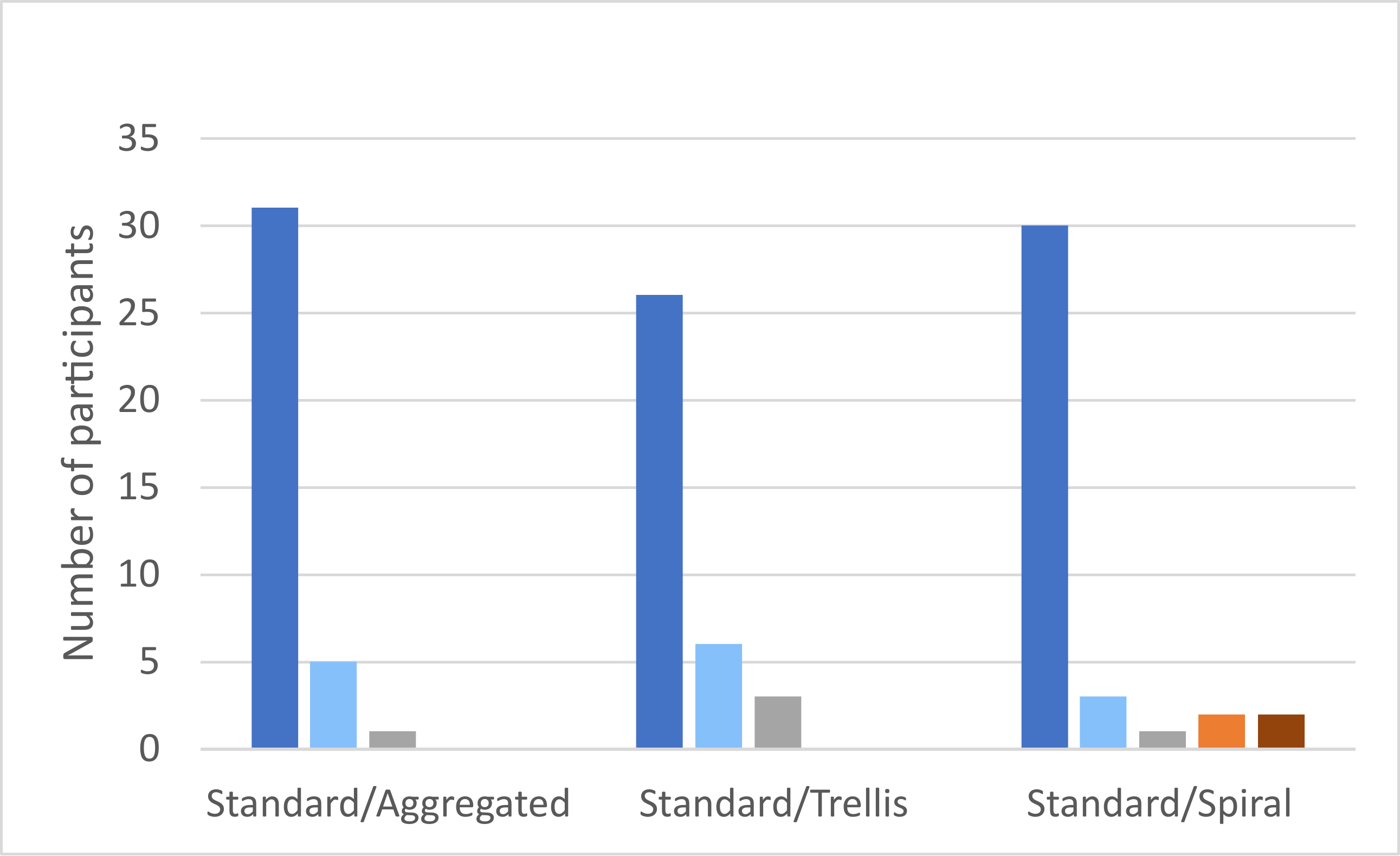}}    
    \subfloat{\includegraphics[width=0.17\linewidth]{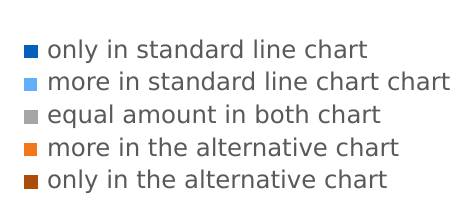}}

    \caption{Participants' response in the trust game for visualization pairs that show the (top) same and (bottom) opposite trends.}
    \label{fig:trust-game-stat}
\end{figure}

\paragraph{Results}
\Cref{fig:trust-game-stat} summarize results for trends.
Numeric values from 1–5 were assigned to investment strategies (1-only in the standard line chart, 5-only in the alternative; 3 = neutral).

\paragraph{Aggregated Charts} \ul{Same Trend Direction:} Up-trend pairs \MeanSdAPA{2.92}{0.649} and down-trend pairs \MeanSdAPA{2.94}{0.791} were near neutral. Slight shifts occurred when one chart’s increase was steeper (toward standard \MeanSdAPA{2.89}{0.522}, toward aggregated \MeanSdAPA{3.06}{0.754}).
\ul{Opposite Trend Direction:} Participants preferred the chart with the higher trend. Low standard deviations indicated consistent responses.

\paragraph{Trellis Charts} \ul{Same Trend Direction:} Up-trend pairs \MeanSdAPA{2.16}{1.056} and steeper-standard cases \MeanSdAPA{2.23}{1.140} showed preference for standard; down-trend pairs were neutral \MeanSdAPA{2.84}{0.986}, and steeper-trellis cases \MeanSdAPA{3.14}{1.134} showed weak preference.
\ul{Opposite Trend Direction:} Participants tended to pick the higher trend, but less decisively than with other chart types (see~\Cref{fig:std_down_same}), suggesting weaker decisiveness and lower trust compared to other chart types.

\paragraph{Spiral Charts} \ul{Same Trend Direction:} Up-trend \MeanSdAPA{3.16}{1.21} and down-trend \MeanSdAPA{3.03}{1.45} pairs were neutral; steeper-standard \MeanSdAPA{3.13}{1.167} and steeper-spiral \MeanSdAPA{3.14}{1.166} cases were also neutral. However, high variance indicated inconsistent responses.
\ul{Opposite Trend Direction:} Participants favored the higher trend. However, this preference weakened in cases where the spiral chart’s upward trend was paired with a steeper downward trend in the standard chart (see~\Cref{fig:std_down_higher}).

\paragraph{Discussion}
Aggregated charts were trusted similarly to standard charts, supporting their use to reduce data density without sacrificing trust. Trellis charts elicited \textbf{lower trust than standard} in most cases and produced less decisive choices in opposite-direction comparisons, except in equal downward trend cases. Spiral charts showed \textbf{mixed trust} with averages near neutral, but high variability suggested individual differences in interpretation.

%% file: sec.discussion_future_work.tex
\begin{wrapfigure}{r}{0.27\linewidth}
          \centering
          \includegraphics[width=0.975\linewidth]{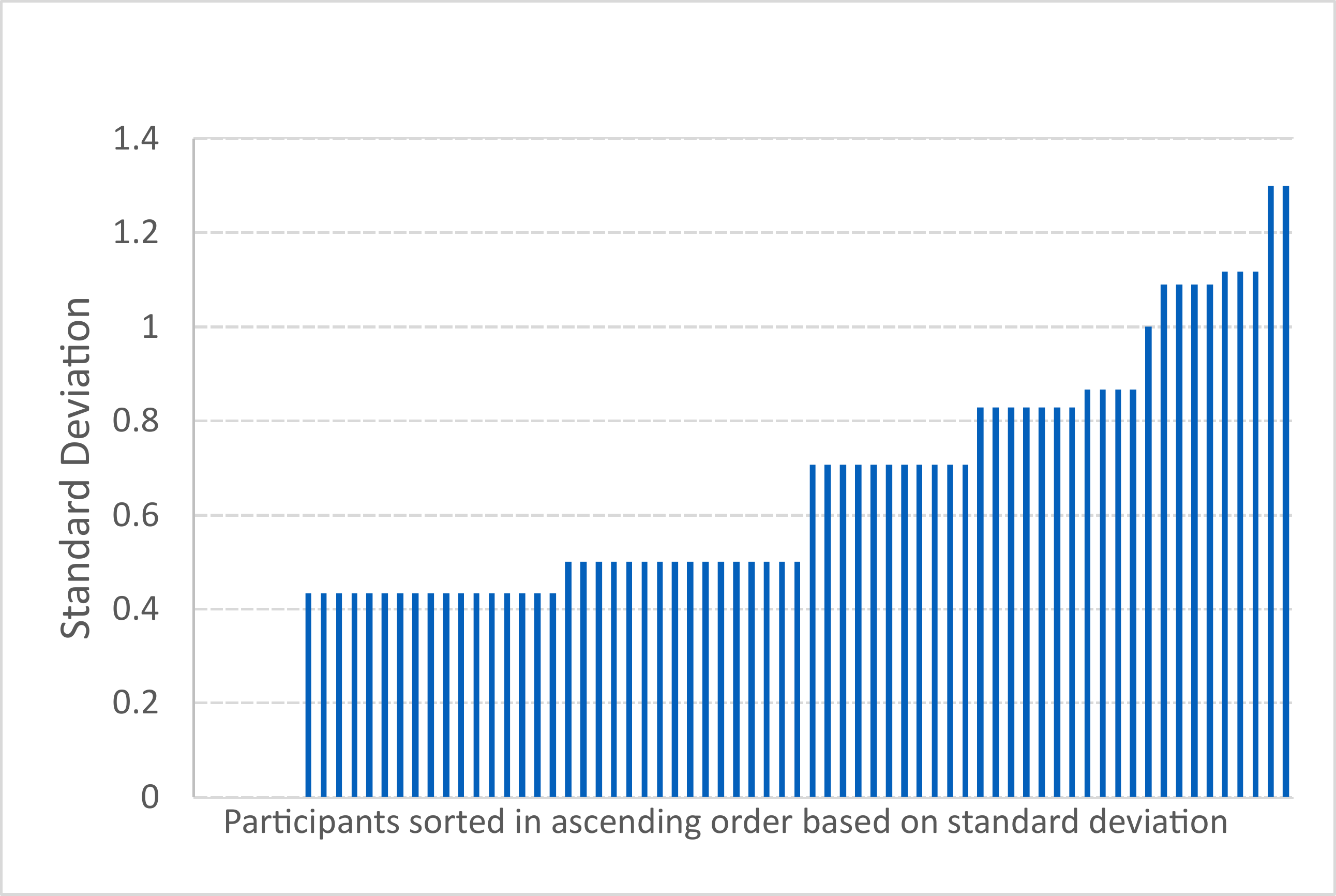}
          \caption{Standard deviation of confidence among participants in the prediction task.}
          \label{fig:prediction-confidence-variance}
    \end{wrapfigure}

\section{Discussion}

We analyzed three strategies for visualizing large time-series data in line charts, namely aggregated charts, trellis charts, and spiral charts, and compared them to a standard line chart in trend, prediction, and decision-making tasks. The key takeaways are:

    \paragraph{Trend Identification in Trellis and Spiral Charts} Standard and aggregated charts more effectively depicted trends than trellis or spiral charts, even for data with periodic patterns. Standard charts consistently outperformed the others, suggesting \textit{these formats should be avoided when trend identification is the goal}.
    
    \paragraph{(Over-)Confidence in Trellis and Spirals} Visualization type did not affect participants' confidence in predictions. On average, they were fairly sure across all charts, yet predictions from trellis and spiral charts diverged significantly from standard charts—\textit{indicating more biased predictions with similar confidence}.

\paragraph{Why Trellis and Spiral Charts Struggled}
     While analyzing the results, we observed that trellis and spiral charts consistently underperformed compared to standard and aggregated line charts. We considered some possible explanations based on their visual design and how participants interacted with them.
    Splitting data by year, \textit{trellis charts} can increase cognitive load, making it harder to perceive overall patterns, especially in data with cyclic trends. As shown in \Cref{fig:line-climate}, periodicity is easier to detect in standard line charts because the entire timeline is visible at once, whereas trellis charts (\Cref{fig:trellis-climate}) require mentally stitching together separate panels.
    On the other hand, many participants were less familiar with \textit{spiral charts}, and their reading order or structure was not always intuitive. Although spirals can fit more data into less space, this compression can obscure periodic patterns, as seen in \Cref{fig:spiral-climate}. The combination of novelty and visual density likely contributed to their lower performance.

    \paragraph{Individual Differences in Confidence} We examined the standard deviation of confidence for individual participants. \Cref{fig:prediction-confidence-variance} indicates that a large number of participants had a high standard deviation in confidence when making predictions using different charts. This implies that participants' confidence levels are affected by various factors, including the visualization approach used. It also suggests that \textit{participants may have different preferences for visualization approaches based on their individual perceptions, experience, and interpretations of the presented data}.

    \paragraph{Trust in Decision-Making} \textit{Aggregated charts emerged as a trustworthy alternative to standard charts}. Trellis and spiral charts were less trusted, raising concerns about their broader use.
    
    \paragraph{Aggregated Charts as a Good Alternative} Our goal was to find a density-reducing method that preserved trend clarity, trust, and predictive tendency. \ul{Results suggest aggregated charts meet these goals well, whereas trellis and spiral charts show notable shortcomings.}

%% file: sec.conclusion.tex
\section{Conclusion}
Our study has some limitations that are important to acknowledge. The datasets used spanned only three years, which constrained the temporal density examined. Longer time spans may increase clutter in standard line charts, potentially giving aggregated charts a greater performance advantage. Additionally, we kept encoding and scaling consistent across all chart types, but design modifications (e.g., redundant color encoding in trellis or spiral charts) could improve their clarity and warrant further investigation.

Overall, our findings indicate that aggregated charts are an effective density-reducing alternative to standard line charts, outperforming trellis and spiral charts in trend detection, prediction accuracy, and decision-making trust. These insights can guide the selection of visualization strategies for clearer and more reliable communication of temporal data. High-resolution figures and user evaluation details are provided in the supplemental material that can be accessed at: \texttt{\small\url{https://osf.io/x9zmr/?view_only=c3a7071971e5420ea5f8b99e5af420dd}}.